\documentclass{article}

\usepackage{arxiv}

\usepackage[utf8]{inputenc} % allow utf-8 input
\usepackage[T1]{fontenc}    % use 8-bit T1 fonts
\usepackage{hyperref}       % hyperlinks
\usepackage{url}            % simple URL typesetting
\usepackage{booktabs}       % professional-quality tables
\usepackage{amsfonts}       % blackboard math symbols
\usepackage{nicefrac}       % compact symbols for 1/2, etc.
\usepackage{microtype}      % microtypography
\usepackage[utf8]{inputenc}
\usepackage{graphicx}
\usepackage{fvextra}
\usepackage{amsmath}
\usepackage{amssymb}
\usepackage{tabularx}
\usepackage{tikz}
\usepackage{float}
\usepackage{listings}
\usepackage[inline]{enumitem}
\usepackage{hyperref}
\usepackage{algpseudocode,algorithm,algorithmicx}
\usetikzlibrary{er,positioning,bayesnet,arrows}

\newcommand{\card}[1]{\left\vert{#1}\right\vert}
\newcommand*\Let[2]{\State #1 $\gets$ #2}
\newcommand*\cp[2]{P(#1 \, | \, #2)}

\algnewcommand\algorithmicforeach{\textbf{for each}}
\algdef{S}[FOR]{ForEach}[1]{\algorithmicforeach\ #1\ \algorithmicdo}

\newcolumntype{Y}{>{\centering\arraybackslash}X}

\lstdefinestyle{mystyle}{
    basicstyle=\ttfamily\footnotesize
}

\title{Selectivity Estimation with Attribute Value Dependencies using Linked Bayesian Networks}

\author{
  Max Halford \\
  IRIT Laboratory \\
  IMT Laboratory \\
  University of Toulouse \\
  \texttt{max.halford@irit.fr} \\
   \And
 Philippe Saint-Pierre \\
  IMT Laboratory\\
  University of Toulouse \\
  \texttt{philippe.saint-pierre@math.univ-toulouse.fr} \\
   \And
 Frank Morvan \\
  IRIT Laboratory \\
  University of Toulouse \\
  \texttt{frank.morvan@irit.fr} \\
}

\begin{document}
\maketitle

\begin{abstract}
Relational query optimisers rely on cost models to choose between different query execution plans. Selectivity estimates are known to be a crucial input to the cost model. In practice, standard selectivity estimation procedures are prone to large errors. This is mostly because they rely on the so-called attribute value independence and join uniformity assumptions. Therefore, multidimensional methods have been proposed to capture dependencies between two or more attributes both within and across relations. However, these methods require a large computational cost which makes them unusable in practice. We propose a method based on Bayesian networks that is able to capture cross-relation attribute value dependencies with little overhead. Our proposal is based on the assumption that dependencies between attributes are preserved when joins are involved. Furthermore, we introduce a parameter for trading between estimation accuracy and computational cost. We validate our work by comparing it with other relevant methods on a large workload derived from the JOB and TPC-DS benchmarks. Our results show that our method is an order of magnitude more efficient than existing methods, whilst maintaining a high level of accuracy.
\end{abstract}

% keywords can be removed

\section{Introduction}

A query optimiser is responsible for providing a good query execution plan (QEP) for incoming database queries. To achieve this, the optimiser relies on a cost model, which tells the optimiser how much a given QEP will cost. The cost model's estimates are in large part based on the selectivity estimates of each operator inside a QEP \cite{ioannidis1996query}. The issue is that selectivity estimation is a difficult task. In practice, huge mistakes are not exceptions but rather the norm \cite{leis2015good}. In turn, this leads the cost model to produce cost estimates that can be wrong by several orders of magnitude \cite{ioannidis1991propagation}. The errors made by the cost model will inevitably result in using QEPs that are far from optimal in terms of memory usage and running time. Moreover, the cost model may also be used by other systems in addition to the query optimiser. For instance, service-level agreement (SLA) negotiation frameworks are based on the assumption that the cost of each query can accurately be estimated by the cost model \cite{yin2018sla}. Cost models are also used for admission control (should the query be run or not?), query scheduling (when to run a query?), progress monitoring (how long will a query?), and system sizing (how many resources should be allocated to run the query?) \cite{wu2013predicting}. Errors made by the cost model may thus have far reaching consequences. Such errors are for the most part due to the inaccuracy of the selectivity estimates.

Selectivity estimates are usually wrong because of the many simplifying assumptions that are made by the cost model. These assumptions are known to be unverified in practice. Nonetheless, they allow the use of simple methods that have a low computational complexity. For example, the \emph{attribute value independence} (AVI) assumption, which states that attributes are independent with each other, is ubiquitous. This justifies the widespread use of one-dimensional histograms for storing the distribution of attribute values. Another assumption which is omnipresent is the \emph{join uniformity assumption}, which states that attributes preserve their distribution when they are part of a join. Although this is a major source of error, it rationalises the use of simple formulas that surmise uniformity \cite{selinger1979access}. Producing accurate selectivity estimates whilst preserving a low computational overhead is thus still an open research problem, even though many methods from various approaches have been proposed.

The standard approach to selectivity estimation is to build a statistical synopsis of the database. The synopsis is built at downtime and is used by the cost model when the query optimiser invokes it. The synopsis is composed of statistics that summarise each relation along with its attributes. Unidimensional constructs, e.g., histograms \cite{ioannidis2003history}, can be used to summarise single attributes, but cannot dependencies between attributes. Multidimensional methods, e.g., multivariate histograms \cite{muralikrishna1988equi}, can be used to summarise the distribution of two or more attributes. However, their spatial requirement grows exponentially with the number of attributes. Moreover, they often require a non-trivial construction phase that takes an inordinate amount of time. Another approach is to use sampling, where the idea is to run a query on a sample of the database and extrapolate the selectivity \cite{chen2017two}. Sampling works very well for single relations. The problem is that sampling is difficult to apply in the case of joins. This is because the join of sampled relations has a high probability of being empty \cite{chaudhuri1999random}. A different approach altogether is to acknowledge that the cost model is mostly wrong, and instead learn from its mistakes so as not to reproduce them. The most successful method in this approach is DB2's so called \emph{learning optimiser} (LEO) \cite{stillger2001leo}. Such a memorising approach can thus be used in conjunction with any cost model. Although they are appealing, memorising approaches do not help in any matter when queries that have not been seen in the past are issued. What's more, they are complementary to other methods. Finally, statistical approaches based on conditional distributions seem to strike the right balance between selectivity estimation accuracy and computational requirements \cite{halford2019approach}. A conditional distribution is a way of modifying a distribution of values based on the knowledge of another value -- called the conditioning value. For example, if the values of attribute $B$ depend on those of $A$, then we can write $P(A, B) = P(B | A) \times P(A)$. Conditional distributions can be organised in a so-called Bayesian network \cite{jensen1996introduction}. Bayesian networks thus factorise a multidimensional distribution into a product of lower dimensional ones. If well chosen, these factorisations can preserve most of the information whilst consuming much less space. In \cite{halford2019approach}, we proposed to use Bayesian networks to capture attribute value dependencies inside each relation of a database. The issue with Bayesian networks is their computational cost \cite{cooper1990computational}. To alleviate this issue, we restricted our networks to possess tree topologies, which leads to simpler algorithms that have the benefit in linear time. The downside of using tree topologies is that our networks capture less dependencies than a general network. However, we showed in our benchmarks that our method was able to improve the overall selectivity estimation accuracy at a very reasonable cost. The downside of our work in \cite{halford2019approach} is that it completely ignores dependencies between attributes of different relations, which we address in the present work.

Bayesian networks that capture attribute value dependencies across relations have also been proposed. \cite{getoor2001selectivity} were the first to apply them for selectivity estimation. However, they used off-the-shelf algorithms that are standard for working with Bayesian networks, but which are costly and impractical in constrained environments. \cite{tzoumas2011lightweight} extended the work of \cite{getoor2001selectivity} to address the computational cost issues. Indeed, they proposed various constraints on the network structure of each relation's Bayesian network that reduced the overall complexity. However, this still results in a global Bayesian network with a complex structure, which requires a costly inference algorithm in order to produce selectivity estimates. Although the methods from \cite{getoor2001selectivity} and \cite{tzoumas2011lightweight} enable a competitive accuracy, they both incur a costly construction phase and are too slow at producing selectivity estimates. In light of this, our goal is to capture attribute value dependencies across relations with as little an overhead as possible. With this in mind, our method thus consists in measuring the distribution of a carefully selected set of attributes before and after a join. We do so by performing a small amount of offline joins that exploits the topology of the relations. Effectively, we make use of the fact that joins mainly occur on primary/foreign key relationships, and thus warp the attribute values distribution in a predictable way. The contributions of our paper are as follows: (1) we introduce a new assumption which simultaneously softens the attribute value independence and join uniformity assumption, (2) based on our assumption, we propose an algorithm for connecting individual Bayesian networks together into what we call a \emph{linked Bayesian network}, (3) we show how such a linked Bayesian network can be used to efficiently estimate query selectivities both within and between relations, and (4) we introduce a parameter which allows us to generalise the trade-offs induced by existing methods based on Bayesian networks.

%For instance, in a one-to-many join only the distributions of the attribute values from the left relation are modified, not those from the right one. By assuming that attribute value dependencies inside a relation are preserved after a join, our procedure is able to induce many new conditional distributions for free, without the large overhead incurred by existing methods \cite{getoor2001selectivity,tzoumas2011lightweight}.

The rest of the paper is organised as follows. Section 2 presents the related work. Section 3 introduces some notions related to Bayesian networks and summarises the work we did in \cite{halford2019approach}. Section 4 introduces the methodology for combining individual Bayesian networks using the topology of a database's relations. In section 5, we compare our proposal with other methods on an extensive workload derived from the JOB \cite{leis2015good} and TPC-DS \cite{poess2002tpc} benchmarks. Finally, section 6 concludes the paper and hints to potential follow-ups.

\section{Related work}

Ever since the seminal work of Selinger et al. \cite{selinger1979access}, query optimisation has largely relied on the use of cost models. Because the most important part of the cost model is the selectivity estimation module \cite{leis2018query}, a lot of good efforts have been made across the decades. \cite{kooi1981optimization} first proposed the use of histograms to approximate the distribution $P(x)$ of a single attribute $x$. Since then, a lot of work has gone into developing optimal histograms \cite{ioannidis2003history} that have been used ubiquitously in cutting edge cost models. Smooth versions of histograms, e.g., kernel density estimators (KDEs) \cite{blohsfeld1999comparison} and wavelets \cite{matias1998wavelet}, have also been proposed. However, these methods are based on single attributes, and as such lose in accuracy what they gain in computational efficiency. Indeed, there is no way to capture a dependency between two attributes $x$ and $y$ if one only has unidimensional distributions $P(x)$ and $P(y)$ available, regardless of their accuracy.

Multidimensional distributions, i.e., $P(X_1, \dots, X_n)$, are a way to catch dependencies between attributes. Methods based on such distributions are naturally more accurate because they soften the AVI assumption. However, they require a large amount of computational resources which hinders their use in high-throughput settings. \cite{muralikrishna1988equi} first formalised the use of equi-depth multidimensional histograms and introduced an efficient construction algorithm. \cite{poosala1997selectivity} proposed another construction algorithm based on Hilbert curves. Multidimensional KDEs have also been proposed \cite{heimel2015self}, with somewhat the same complexity guarantees. In search for efficiency, \cite{bruno2001stholes} offered a workload-aware method where the idea is to only build histograms for attributes that are often queried together. Even though methods based on multidimensional distributions are costly, they are implemented in some database systems and are used when specified by a database administrator. However, these methods do not help whatsoever in capturing dependencies across relations, which is probably the biggest issue cost models have to deal with.

Sampling methods have also been proposed to perform selectivity estimation. The idea is to run a query on a sample of the database and extrapolate the selectivity \cite{chen2017two}. Sampling works very well for single relations and has been adopted by some commercial database systems. However, off the shelf sampling procedures suffer from the fact that the join of sampled relations has a high probability of being empty \cite{chaudhuri1999random}; in other words a join has to be materialised before sampling can be done. This issue can be alleviated by the use of correlated sampling \cite{vengerov2015join}, where a deterministic hash function is used to ensure that samples from different relations will match with each other. Another technique is to use indexes when available \cite{leis2017cardinality}, but this is only realistic for in-memory databases. \cite{acharya1999join} also proposed heuristics for maintaining statistics of \emph{join synopses}. Overall, sampling is an elegant selectivity estimation method, not least because it can handle complex predicates which statistical summaries cannot (e.g., regex queries). However, sampling necessarily incurs a high computational cost. Indeed even if the samples are obtained at downtime, they still have to be loaded in memory during the query optimisation phase.

Throughout the years, a lot of proposals have been made to relax the simplifying assumptions from \cite{selinger1979access}. All of these require compromises in terms of accuracy, speed, and memory usage. The general consensus is that each method shines in a particular use case, and thus combining different methods might be a good approach. \cite{markl2007consistent} formalised this idea by using a maximum entropy approach. Recently, \cite{muller2018improved} proposed combining sampling and synopses. Another approach altogether is to ``give up'' on the cost model and instead memorise the worst mistakes it makes so as not to reproduce them in the future \cite{stillger2001leo}. There have also been proposals that make use of machine learning \cite{akdere2012learning,liu2015cardinality,ivanov2017adaptive,kipf2018learned}, where a \emph{supervised learning} algorithm is taught to predict selectivities based on features derived from a given query and the database's metadata. Recently, deep learning methods have been proposed to extract features that don't require rules written by humans. One of the most prominent papers that advocates the use of deep learning for selectivity estimation can be found in \cite{kipf2019estimating}. They proposed a neural network architecture, which they dubbed MSCN for \emph{multi-set convolutional network}. Although approaches based on supervised machine learning have had great success in other domains, their performance for query selectivity estimation isn't competitive enough, yet.

Approaches that exploit attribute correlations in order to avoid storing redundant information have also been proposed. For example, \cite{deshpande2001independence} proposes to build a statistical interaction model that allows to determine a relevant subset of multidimensional histograms to build. In other words, they propose to build histograms when attributes are correlated, and to make the AVI assumption if not. Bayesian networks can be seen through the same lens of exploiting redundant information. Essentially, they factorise the full probability distribution into a set of conditional distributions. A conditional distribution between two attributes implies a hierarchy whereby one of the attributes determines to some extent the other. Formally, a Bayesian network is a directed acyclic graph (DAG) where each node is an attribute and each arrow implies a conditional dependency. They can be used to summarise a probability distribution by breaking it up into smaller pieces. In comparison with the supervised learning based methods mentioned in the previous paragraph, Bayesian networks are an \emph{unsupervised learning} method. What this means is that they directly learn by looking at the data, whereas supervised methods require a workload of queries and outputs in order to learn. In \cite{halford2019approach}, we proposed to use Bayesian networks for capturing attribute value dependencies inside individual relations. \cite{getoor2001selectivity} and \cite{tzoumas2011lightweight} both proposed methods for using Bayesian networks to capture attribute value dependencies between different relations. Although this leads to more accurate selectivity estimates, it requires much more computation time and is infeasible in practice. This is due to the fact that they require the use of expensive belief propagation algorithms for performing inference. Meanwhile, our method is much faster because it restricts each Bayesian network to a tree topology, which allows the use of the variable elimination algorithm. However, our method completely ignores dependencies between attributes of different relations. Our goal in this paper is to reconcile both approaches. Effectively, we want to keep the computational benefits of building and using individual Bayesian networks, but at the same time we want our method to capture some dependencies across relations.

\section{Methodology}

\subsection{Preliminary work}

In \cite{halford2019approach}, we developed a methodology for constructing Bayesian networks to model the distribution of attribute values inside each relation of a database. A Bayesian network is a probabilistic model. As such, it is used for approximating the probability distribution of a dataset. The particularity of a Bayesian network is that it uses a directed acyclic graph (DAG) in order to do so. The graph contains one node per variable, whilst each directed edge represents a conditional dependency between two variables. Therefore, the graph is a factorisation of the full joint distribution:

\begin{equation}
    P(X_1, \dots, X_n) \simeq \prod_{X_i \in \mathcal{X}} P(X_i \, | \, Parents(X_i))
\end{equation}

The joint distribution $P(X_1, \dots, X_n)$ is the probability distribution over the entire set of attributes $\{X_1, \dots, X_n\}$. Meanwhile, $Parents(X_i)$ stands for the attributes that condition the value of $X_i$. The distribution $P(X_i \, | \, Parents(X_i))$ is thus the conditional distribution of attribute $X_i$'s value. In practice, the full distribution is inordinately large, and is unknown to us. However, the total of the sizes of the conditional distributions $P(X_i \, | \, Parents(X_i))$ is much smaller. 

Using standard rules of probability, such as Bayes' rule and the law of total probability \cite{jensen1996introduction}, we are able to derive from a Bayesian network any selectivity estimation problem by converting a logical query into a product of conditional probabilities. Note, however, that a Bayesian network is necessarily an approximation of the full probability distribution because it makes assumptions about the generating process of the data. Finding the right graph structure of a Bayesian network is called \emph{structure learning} \cite{jensen1996introduction}.

This is usually done by maximising a scoring function, which is an expensive process that scales super-exponentially with the number of variables \cite{cooper1990computational}. Approximate search methods as well as integer programming solutions have been proposed \cite{bartlett2017integer}. In our work in \cite{halford2019approach}, we proposed to use the \emph{Chow-Liu} algorithm \cite{chow1968approximating}. This algorithm has the property of finding the best tree structure where nodes are restricted to have at most one parent. The obtained tree is the best in the sense of maximum likelihood estimation. In addition to this property, the Chow-Liu algorithm only runs in $\mathcal{O}(p^2)$ time, where $p$ is the number of variables, and is simple to implement. It works by first computing the \emph{mutual information} between each pair of variables, which can be seen as the strength of the relation between two variables.

The next step is to find the \emph{maximum spanning tree} (MST) using the mutual information, and thus to derive a directed graph approximating the joint probability distribution. We propose an inference process based on variable elimination algorithm \cite{cowell2006probabilistic} since inference can be done in linear time for tree. Our experiments indicated that competitors approach are much slower. Note that our inference process can further be accelerated using the Steiner tree problem \cite{hwang1992steiner}.

In \cite{halford2019approach}, we proposed a simple method which consists in building one Bayesian network per relation. On the one hand, this has the benefit of greatly reducing the computational burden in comparison with a single large Bayesian network, as is done in \cite{getoor2001selectivity} and \cite{tzoumas2011lightweight}. On the other hand, it ignores dependencies between attributes of different relations. We will now discuss how we can improve our work from \cite{halford2019approach} in order to capture some dependencies across relations.

\subsection{Handling conditional dependencies over joins}

The task of selectivity estimation is to determine the selectivity of a query over a set of attributes $X_i$ that to a set of relations $R_j$. By making the AVI assumption, this comes down to to measuring individual attribute distributions and multiplying them together, as so:

\begin{equation} \label{infer-ibn}
    P(X_1, \dots, X_n) \simeq \prod_{R_j} \bigg( \prod_{X_i \in R_j} P(X_i) \bigg)
\end{equation}

The methodology from \cite{halford2019approach} models the attribute distribution value of a database by building a tree-shaped Bayesian network for each relation. For efficiency reasons it purposefully only captures dependencies between attributes of a single relation. As such, it ignores the many dependencies that exist between attributes of different relations and that are the bane of cost models. Essentially, this method boils down to factorising the full probability distribution as so:

\begin{equation}
    P(X_1, \dots, X_n) \simeq \prod_{R_j} \bigg( \prod_{X_i \in R_j} P(X_i \, | \, Parent(X_i)) \bigg)
\end{equation}

where $\{X_1, \dots, X_n\}$ is the entire set of attributes over all relations and $X_i \in R_j$ are the attributes that belong to relation $R_j$. $Parent(X_i)$ denotes the attribute on which the distribution of $X_i$ is conditioned -- because each Bayesian network is tree-shaped, each attribute excluding the root has a single parent. Although the work from \cite{halford2019approach} ignores dependencies between attributes of different relations, it is still much more relevant that the common assumption of full attribute value independence. Our goal in this paper is to model the full probability distribution by taking into account dependencies between attributes of different relations, which can represented as:

\begin{equation} \label{infer-lbn}
    P(X_1, \dots, X_n) \simeq \prod_{X_i} P(X_i \, | \, Parent(X_i))
\end{equation}

Note that equation \ref{infer-lbn} captures more information than equation \ref{infer-ibn}. Modeling the data by taking into account conditional dependencies thus guarantees that the resulting selectivity estimates are at least as accurate as when assuming independence between attributes. In \cite{halford2019approach}, we made the assumption that attribute values of different relations are independent. Additionally, we assumed that each attribute value distribution remains the same when the relation it belongs to is joined with another relation. This is called the \emph{join uniformity assumption} and is a huge source of error. Indeed, the distributions of an attribute's values before and after a join are not necessarily the same. For instance, imagine an e-commerce website where registered customers are stored in a database alongside with purchases. Each customer can make zero or more purchases whilst each purchase is made by exactly one customer. Some customers might be registered on the website but might not have made a purchase. If the customers and purchases relations are joined together, then the customers who have not made any purchase will not be included in the join. Therefore, the attributes from the customers relation will have different value distributions when joined with the purchases relation. Note however that the attribute value distributions from the purchases relation will not be modified. This stems from the fact that the join between the customers and purchases relations is a one-to-many join. We will now explain how we can use this property to capture attribute value dependencies across relations. 

Let us assume we have two relations $R$ and $S$ that share a primary/foreign key relationship. That is, $S$ contains a foreign key which references the primary key of $R$. This means that each tuple from $R$ can be joined with zero or more tuples from $S$. A direct consequence is that the size of the join $R \bowtie S$ is equal to $\card{S}$. The join uniformity assumptions implies that the probability for a tuple $r$ from relation $R$ to be present in $R \bowtie S$ follows a uniform distribution. In statistical terms, that is:

\begin{equation}
    P(r \in R \bowtie S) \sim \mathcal{U}(\frac{1}{\card{R}})
\end{equation}

Consequently, the expected number of times each tuple from $R$ will be part of $R \bowtie S$ is $\frac{\card{S}}{\card{R}}$. Let us now denote by $P_{R}(A)$ the value distribution of attribute $A$ in relation $R$. We will also define $P_{R \bowtie S}(A)$ as the value distribution of attribute $A$ in join $R \bowtie S$. The join uniformity assumption thus implies that the distribution of $A$'s values before and after $R$ is joined with $S$ are equal:

\begin{equation}
    P_{R}(A) = P_{R \bowtie S}(A)
\end{equation}

Furthermore, assume we have found and built the following factorised distribution over attributes $A$, $B$, and $C$ from relation $R$:

\begin{equation}
P_{R}(A, B, C) \simeq P_{R}(A \, | \, B) \times P_{R}(B \, | \, C) \times P_R(C)
\end{equation}

If we hold the join uniformity assumption to be true, then we can use the factorised distribution to estimate selectivities for queries involving $A$, $B$ and $C$ when $R$ is joined with $S$ without any further modification. The issue is that this is an idealised situation that has no reason to occur in practice. On the contrary, it is likely are that some tuples from $R$ will be more or less present than others. However, we may assume that after the join the attribute value dependencies implied by our factorisation remain valid within each relation. We call this the \emph{attribute value dependency preservation} assumption. The idea is that if attributes $A$, $B$ and $C$ are dependent in a certain way in relation $R$, then there is not much reason to believe that these dependencies will disappear once $R$ is joined with $S$. Although this may not necessarily always occur in practice, it is still a much softer assumption than those usually made by cost models.

To illustrate, let us consider a toy database composed of the following relations: \textcolor{teal}{customers} with attributes $\{nationality, hair, salary\}$, \textcolor{orange}{shops} with attributes $\{name, city, size\}$, \textcolor{violet}{purchases} with attributes $\{day\ of\ week\}$. Moreover, assume that the \textcolor{violet}{purchases} relation has two foreign keys, one that references the primary key of \textcolor{teal}{customers} and another which that of \textcolor{orange}{shops}. The \textcolor{violet}{purchases} relation can thus be seen as a fact table whilst \textcolor{teal}{customers} and \textcolor{orange}{shops} can be viewed as dimension tables. represented in table. In what follows we will use the shorthand \textcolor{teal}{$C$} the \textcolor{teal}{customers} relation, \textcolor{orange}{$S$} for the \textcolor{orange}{shops} relation, and \textcolor{violet}{$P$} for the \textcolor{violet}{purchases} relation.

In the \textcolor{teal}{customers} relation, there are Swedish customers and a lot of them have blond hair. We might capture this property in a Bayesian network with the conditional distribution $P_{\textcolor{teal}{C}}(hair \, | \, nationality)$, which indicates that hair colour is influenced by nationality. We could suppose that the fact that Swedish people have blond hair is still true once the \textcolor{teal}{customers} relation is joined with the \textcolor{violet}{purchases} relation. In other words, the hair colour shouldn't change the rate at which Swedish customers make purchases. However, we may rightly assume that the number of purchases will change according to the nationality of each customer. Mathematically, we are saying the following:

\begin{equation}
P_{\textcolor{teal}{C} \bowtie \textcolor{violet}{P}}(hair, nationality) = P_{\textcolor{teal}{C}}(hair \, | \, nationality) \times P_{\textcolor{teal}{C} \bowtie \textcolor{violet}{P}}(nationality)
\end{equation}

In other words, because we assume that $P_{\textcolor{teal}{C}}(hair \, | \, nationality)$ is equal to $P_{\textcolor{teal}{C} \bowtie \textcolor{violet}{P}}(hair \, | \, nationality)$, then we know $P_{\textcolor{teal}{C} \bowtie \textcolor{violet}{P}}(hair, nationality)$  -- i.e.,\ we assume that their conditional distribution remains unchanged after the join. An immediate consequence is that we get to know the $P_{\textcolor{teal}{C} \bowtie \textcolor{violet}{P}}(hair)$ distribution for free. Indeed, by summing over the nationalities, we obtain:

\begin{equation}
P_{\textcolor{teal}{C} \bowtie \textcolor{violet}{P}}(hair) = \sum_{nationality} P_{\textcolor{teal}{C}}(hair \, | \, nationality) \times P_{\textcolor{teal}{C} \bowtie \textcolor{violet}{P}}(nationality)
\end{equation}

To demonstrate why our assumption is useful for the purpose of selectivity estimation, let us use the example data in tables \ref{tab:customers} and \ref{tab:purchases}.

\begin{table}[!htb]
\begin{minipage}{.5\linewidth}
      \centering
        \caption{\textcolor{teal}{Customers} relation}
         \label{tab:customers}
        \begin{tabularx}{\textwidth}{@{}YYY@{}}
            \textbf{Customer} & \textbf{Nationality}  & \textbf{Hair} \\
 \hline
    1  & Swedish     & Blond \\
    2  & Swedish     & Blond \\ 
    3  & Swedish     & Brown \\ 
    4  & American    & Blond \\
    5  & American    & Brown 
        \end{tabularx}
    \end{minipage} 
    \begin{minipage}{.5\linewidth}
      \caption{\textcolor{violet}{Purchases} relation, which contains a foreign key that is related to the primary key of the customers relation}
      \label{tab:purchases}
      \centering
       \begin{tabularx}{0.6\textwidth}{@{}YY@{}}
           \textbf{Shop} & \textbf{Customer} \\
 \hline
    1  & 1 \\
    2  & 1 \\
    3  & 1 \\
    4  & 1 \\
    5  & 2 \\
    6  & 3 \\
    7  & 5 
        \end{tabularx}
    \end{minipage}%
\end{table}

Let us say we wish to know how many purchases are made by customers who are both blond and Swedish. The straightforward way to do this is to count the number of times ``Blond'' and ``Swedish'' appear together within the join $C \bowtie P$:

\begin{equation}
P_{\textcolor{teal}{C} \bowtie \textcolor{violet}{P}}(hair=Blond, nationality=Swedish) = \frac{5}{7}
\end{equation}

The fraction $\frac{5}{7}$ is the true amount of purchases that were made by Swedish customers with blond hair -- said otherwise this is the selectivity of the query. Obtaining it requires scanning the rows resulting from the join of \textcolor{teal}{$C$} with \textcolor{violet}{$P$}. In practice this can be very burdensome, especially when queries involve many relations. If we assume that the join uniformity assumption holds -- in other words we assume that the value distributions of \emph{nationality} and \emph{hair} do not change -- then we can simply reuse the Bayesian network of the customers relation:

\begin{equation}
\begin{split}
    P_{\textcolor{teal}{C} \bowtie \textcolor{violet}{P}}(Blond, Swedish) & \simeq P_{\textcolor{teal}{C}}(Blond \, | \, Swedish) \times P_{\textcolor{teal}{C}}(Swedish) \\
    & \simeq \frac{2}{3} \times \frac{3}{5} \\
    & \simeq \frac{2}{5}
\end{split}
\end{equation}

In this case, making the join uniformity independence assumption makes us underestimate the true selectivity by 44\% ($1 - \frac{2}{5} \times \frac{7}{5}$). Some of this error is due to the fact that the nationality attribute values are not distributed in the same way once \textcolor{teal}{$C$} and \textcolor{violet}{$P$} are joined -- indeed in this toy example Swedish customers make more purchases than American ones. However, if we know the distribution of the nationality attribute values, i.e.,\ $P_{\textcolor{teal}{C} \bowtie \textcolor{violet}{P}}(nationality)$, then we can enhance our estimate in the following manner:

\begin{equation}
\begin{split}
    P_{\textcolor{teal}{C} \bowtie \textcolor{violet}{P}}(Blond, Swedish) & \simeq P_{\textcolor{teal}{C}}(Blond \, | \, Swedish) \times P_{\textcolor{teal}{C} \bowtie P}(Swedish) \\
    & \simeq \frac{2}{3} \times \frac{6}{7} \\
    & \simeq \frac{4}{7}
\end{split}
\end{equation}

Now our underestimate has shrunk to 20\%. The only difference with the previous equation is that we have replaced $P_{\textcolor{teal}{C}}(Swedish)$ with $P_{\textcolor{teal}{C} \bowtie \textcolor{violet}{P}}(Swedish)$. Note that we did not have to precompute $P_{\textcolor{teal}{C} \bowtie \textcolor{violet}{P}}(Blond, Swedish)$. Indeed, we assumed that the dependency between nationality and hair doesn't change once \textcolor{teal}{$C$} and \textcolor{violet}{$P$} are joined, which stems from our dependency preservation assumption. Note that, in our toy example, the assumption is slightly wrong because blond customers have a higher purchase rate than brown haired ones, regardless of the nationality. Regardless, our assumption is still much softer than the join uniformity and attribute value independence assumptions.

Our assumption is softer than the join uniformity assumption because it allows attribute value distributions to change after a join. Statistically speaking, instead of assuming that tuples appear in a join following a uniform distribution, we are saying that the distribution of the tuples is conditioned on a particular attribute (e.g., the nationality of the customers dictates the distribution of the customers in the join between shops and customers). We also assume that attribute value dependencies with each relation are preserved through joins (e.g., hair colour is still dependent on nationality). The insight is that in a factorised distribution, the top-most attribute is part of any query. For instance, in the distribution $\cp{A}{B} \times \cp{B}{C} \times P(C)$, every query involving any combination of $A$, $B$, and $C$ will necessarily involve $P(C)$. We will now see how our newly introduced attribute value dependency preservation assumption can be used to link Bayesian networks from different relations together, and as such relax the join uniformity and attribute value independence assumptions at the same time.

\subsection{Linking Bayesian networks} \label{lbn}

As explained in the previous subsection, if a \textcolor{violet}{purchases} relation has a foreign key that references a primary key of another relation named \textcolor{teal}{customers}, then the distribution of \textcolor{violet}{purchases}' attribute values will not change after joining \textcolor{teal}{customers} and \textcolor{violet}{purchases}. However the distribution of \textcolor{teal}{customers}' attribute values will change if \textcolor{violet}{purchases}' foreign key is skewed, which is always the case to some degree. If we use the method proposed by \cite{halford2019approach}, then the Bayesian network built on \textcolor{teal}{customers} would not be accurate when estimating selectivities for queries involving \textcolor{teal}{customers} and \textcolor{violet}{purchases}. This is because it would assume the distributions of the attribute values from \textcolor{teal}{customers} are preserved after the join, which is a consequence of the join uniformity assumption. Moreover, because of the AVI assumption, we would not be capturing the existing dependencies between \textcolor{teal}{customers}'s attributes and \textcolor{violet}{purchases}'s attributes because their respective attributes are assumed to be independent with those of the opposite relation. On the other hand, if we join \textcolor{teal}{customers} and \textcolor{violet}{purchases} and build a Bayesian network on top of the join, then we will capture the cross-relation attribute value dependencies, but at too high a computational cost \cite{getoor2001selectivity,tzoumas2011lightweight}. Up to now, we have only mentioned the case where there one join occurs, but the same kind of issues occur for many-way joins -- including star-joins and chain-joins.

If the attribute value distributions of \textcolor{teal}{customers} and \textcolor{violet}{purchases} are estimated using Bayesian networks that possess a tree structure, then we only have to include the dependencies of a subset of\textcolor{teal}{customers}'s attributes with those of \textcolor{violet}{purchases}. Specifically, we only have to include the root attribute of \textcolor{teal}{customers}'s Bayesian network into that of of \textcolor{violet}{purchases}. Indeed, because \textcolor{teal}{customers}'s Bayesian network is a tree, then all of its nodes are necessarily linked to the root. If we know the distribution of the root attribute's values \emph{after} \textcolor{teal}{customers} is joined with \textcolor{violet}{purchases}, then, by making the attribute value dependency preservation assumption earlier introduced, we automatically obtain the distribution of the rest of \textcolor{teal}{customers}'s attribute. In other words, if the distribution of an attribute's values is modified when the relation it belongs to is joined with another relation, then we assume that all the attributes that depend on it have their value distributions modified in the exact same manner. This is another way of saying that the conditional distributions remain the same.

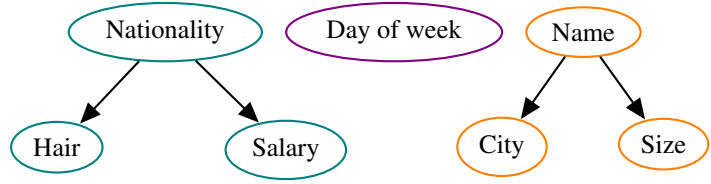
\begin{figure}[H]
  \centering
        \begin{tikzpicture}[thick, scale=\textwidth, every node/.style={ellipse, draw, align=center},node distance=1cm and 0.1cm]

            \node [draw=violet] (dow) {Day of week};
            
            \node [left = 0.3 of dow, draw=teal] (nationality) {Nationality};
            \node [below left = of nationality, draw=teal] (hair) {Hair};
            \node [below right = of nationality, draw=teal] (salary) {Salary};
            
            \node [right = 0.3 of dow, draw=orange] (name) {Name};
            \node [below left = of name, draw=orange] (city) {City};
            \node [below right = of name, draw=orange] (size) {Size};
            
            \draw [->] (nationality) -- (hair);
            \draw [->] (nationality) -- (salary);
            
            \draw [->] (name) -- (city);
            \draw [->] (name) -- (size);
        
        \end{tikzpicture}
\caption{Separate Bayesian networks of \textcolor{teal}{customers}, \textcolor{orange}{shops}, and \textcolor{violet}{purchases}}
\label{fig:separate-network-topology}
\end{figure}

We will show how this works on our toy database consisting of relations \textcolor{teal}{customers}, \textcolor{orange}{shops}, and \textcolor{violet}{purchases}. Following the methodology from \cite{halford2019approach}, we would have built one Bayesian network per relation. Each Bayesian network would necessarily have been a tree as a consequence of using the Chow-Liu algorithm \cite{chow1968approximating}. Depending on the specifics of the data, we might have obtained the Bayesian networks shown in figure \ref{fig:separate-network-topology}.

Furthermore, let us consider the following SQL query:

\begin{figure}[H]
    \centering
\begin{lstlisting}[language=SQL]
    SELECT *
    FROM customers, shops, purchases
    WHERE customers.id = purchases.customer_id
    AND shops.id = purchases.shop_id
    AND customers.nationality = 'Japanese'
    AND customers.hair = 'Dark'
    AND shops.name = 'Izumi'
\end{lstlisting}
\end{figure}

If we were to estimate the amount of tuples that satisfy the above query using the Bayesian networks from figure \ref{fig:separate-network-topology}, then we would estimate the query selectivity in the following manner:

\begin{equation}
    \begin{split}
    P(Dark, Japanese, Izumi) & = P_{\color{teal}{C}}(Dark \, | \, Japanese) \\
                                 & \times P_{\color{teal}{C}}(Japanese) \\
                                 & \times P_{\textcolor{orange}{S}}(Izumi)
    \end{split}
\end{equation}

\begin{figure}[H]
  \centering
        \begin{tikzpicture}[thick, scale=\textwidth, every node/.style={ellipse, draw, align=center},node distance=1cm and 0.1cm]

            \node [draw=violet] (p_nationality) {Nationality};
            \node [below = of p_nationality, draw=violet] (dow) {Day of week};
            \node [below right = of p_nationality, xshift=1cm, draw=violet] (p_name) {Name};
            
            \node [below left = of p_nationality, xshift=-1cm, draw=teal] (nationality) {Nationality};
            \node [below left = of nationality, draw=teal] (hair) {Hair};
            \node [below right = of nationality, draw=teal] (salary) {Salary};
            
            \node [below = 0.5 of p_name, draw=orange] (name) {Name};
            \node [below left = of name, draw=orange] (city) {City};
            \node [below right = of name, draw=orange] (size) {Size};
            
            \draw [->] (p_nationality) -- (dow);
            \draw [->] (p_nationality) -- (p_name);

            \draw [dotted] (p_nationality) -- (nationality);
            
            \draw [->] (nationality) -- (hair);
            \draw [->] (nationality) -- (salary);
            
            \draw [dotted] (p_name) -- (name);
            
            \draw [->] (name) -- (city);
            \draw [->] (name) -- (size);
        
        \end{tikzpicture}
\caption{Linked Bayesian network of \textcolor{teal}{customers}, \textcolor{orange}{shops}, and \textcolor{violet}{purchases}}
\label{fig:linked-network-topology}
\end{figure}
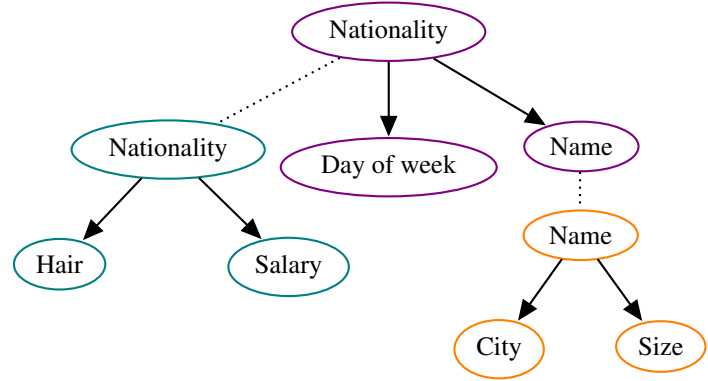

On the one hand, the conditional distribution $P_{\color{teal}{C}}(Dark \, | \, Japanese)$ captures the fact that Japanese people tend to have dark hair inside the \textcolor{teal}{customers} relation. Graphically this is represented by the arrow that points from the ``Nationality'' node to the ``Hair'' node in figure \ref{fig:separate-network-topology}. On the other hand, our estimate ignores the fact that shops in Japan, including ``Izumi'', are mostly frequented by Japanese people. The reason why is that we have one Bayesian network per relation, instead of a global network spanning all relations, and are thus not able to capture this dependency. Regardless of the missed dependency, this simple method is still more accurate than assuming total independence. Indeed the AVI assumption would neglect the dependency between $hair$ and $nationality$, even though both attributes are part of the same relation. Meanwhile assuming relational independence is convenient because it only requires capturing dependencies within relations, but it discards the dependency between $nationality$ and $city$. We propose to capture said dependency by adding nodes from the Bayesian networks of \textcolor{teal}{customers} and \textcolor{orange}{shops} to the Bayesian network of \textcolor{violet}{purchases}. Specifically, for reasons that will become clear further on, we add the roots of the Bayesian networks of \textcolor{teal}{customers} and \textcolor{orange}{shops} (i.e., $nationality$ and $name$) to the Bayesian network of \textcolor{violet}{purchases}. This results in the linked Bayesian network shown in figure \ref{fig:linked-network-topology}.

In this new configuration, we still have one Bayesian network per relation. The difference is that the Bayesian network of \textcolor{violet}{purchases} includes the root attributes of both \textcolor{teal}{customers} and \textcolor{orange}{shops}'s Bayesian networks. In other words, we have joined the \textcolor{violet}{purchases} relation with the \textcolor{teal}{customers} and \textcolor{orange}{shops} and we have then built a Bayesian network for \textcolor{violet}{purchases} that now includes attributes from \textcolor{teal}{customers} and \textcolor{orange}{shops}. A linked Bayesian network is thus a set of separate Bayesian networks where some of the attributes are duplicated in two related networks. In practice, this means that we now know the distribution of the $nationality$ and $name$ attribute values once the relations they belong to have been joined with \textcolor{violet}{purchases}. Meanwhile, we also know their distributions when these relations are not joined with \textcolor{violet}{purchases}. In other words, we store two distributions for each root attribute, one before the join and one afterwards. The distribution of a root attribute in a Bayesian network is nothing more than a one-dimensional histogram. This means that storing two distributions for each root attribute doesn't incur any significant memory burden.

The configuration shown in figure \ref{fig:linked-network-topology} has two immediate benefits over the one presented in figure \ref{fig:separate-network-topology}. First of all, we are now able to determine if the percentage of Japanese in the \textcolor{violet}{purchases} relation is different from the one in the \textcolor{teal}{customers} relation. Indeed, we do not have to assume the distribution remains the same after the join now that we know the distribution of $nationality$'s values when \textcolor{teal}{customers} is joined with \textcolor{violet}{purchases}. A key observation is that we get to know something about the distribution of the $hair$ attribute values when \textcolor{teal}{customers} is joined with \textcolor{violet}{purchases}. That is to say, because we know how the distribution of $nationality$ attribute values changes after the join, then we also know something about the $hair$ attribute values because both attributes are dependent within the \textcolor{teal}{customers} relation. This stems from the fact that we assume that the conditional distribution $\cp{hair}{nationality}$ is preserved after the join. Mathematically this translates to:

\begin{equation}
    P_{ \textcolor{teal}{C} \bowtie \textcolor{violet}{P}}(hair, nationality) = P_{\textcolor{teal}{C}}(hair \, | \, nationality) P_{\textcolor{teal}{C} \bowtie \textcolor{violet}{P}}(nationality)
\end{equation}

Although, in practice, we expect the dependency preservation assumption to not always be verified, we argue that it is a much weaker assumption than assuming total relational independence. The second benefit is that we can now take into account the fact the Japanese people typically shop in Japanese shops, even though the involved attributes belong to relations that are not directly related. This happens because the $name$ attribute is now part of \textcolor{violet}{purchases}'s Bayesian network as well as that of \textcolor{orange}{shops}. Formally the query selectivity can now be expressed as so:

\begin{equation}
    \begin{split}
    P_{\textcolor{teal}{C} \bowtie \textcolor{violet}{P} \bowtie \textcolor{orange}{S}}(Dark, Japanese, Izumi) & = P_{\color{teal}{C}}(Dark \, | \, Japanese) \\
                                 & \times P_{\textcolor{violet}{P} \bowtie \textcolor{orange}{S}}(Izumi \, | \, Japanese) \\
                                 & \times P_{\textcolor{teal}{C} \bowtie \textcolor{violet}{P}}(Japanese)
    \end{split}
\end{equation}

Let us now consider the following SQL query where the only difference with the previous query is that are filtering by $city$ instead of by $name$:

\begin{figure}[H]
    \centering
    \begin{lstlisting}[language=SQL]
        SELECT *
        FROM customers, shops, purchases
        WHERE customers.id = purchases.customer_id
        AND shops.id = purchases.shop_id
        AND customers.nationality = 'Japanese'
        AND customers.hair = 'Dark'
        AND shops.city = 'Osaka'
    \end{lstlisting}
\end{figure}

In this case, our linked Bayesian network would estimate the selectivity as so:

\begin{equation}
    \begin{split}
    P(Dark, Japanese, Osaka) & = P_{\color{teal}{C}}(Dark \, | \, Japanese) \\
                                 & \times \sum_{name}
                                    P_{\textcolor{violet}{P} \bowtie \textcolor{orange}{S}}(Osaka \, | \, name) P_{\textcolor{violet}{P}}(name \, | \, Japanese) \\
                                & \times P_{\textcolor{teal}{C} \bowtie \textcolor{violet}{P}}(Japanese)
    \end{split}
\end{equation}

This is a simple application of Bayesian network arithmetic \cite{jensen1996introduction}. The reason why there is a sum is that we have to take into account all the shops that are located in Osaka because none of them in particular has been specified in the SQL query. Note that our linked Bayesian network is still capable of estimating selectivities when only a single relation is involved. For example, we only need to use $P_{\textcolor{violet}{P}}(nationality)$ when the \textcolor{teal}{customers} relation is joined with \textcolor{violet}{purchases} relation. If only the \textcolor{teal}{customers} relation is involved in a query, then we can simply use $P_{\textcolor{violet}{C}}(nationality)$ instead of $P_{\textcolor{violet}{P}}(nationality)$. We discuss these two points in further detail in subsection \ref{querying}.

Linked Bayesian networks thus combine the benefits of independent Bayesian networks, while having the benefit of softening the join uniformity assumption as well as the attribute value independence assumption. We will now discuss how one may obtain a linked Bayesian network in an efficient manner.

\subsection{Building linked Bayesian networks} \label{building}

A linked Bayesian network is essentially a set of Bayesian networks. Indeed, our method consists in taking individual Bayesian networks and linking them together in order to obtain one single Bayesian network. This linking process is detailed in the next subsection. In our case, by only including the root attribute of each relation into the Bayesian network of its parent relation, we ensure that the final network necessarily has a tree topology. Performing inference on a Bayesian network with a tree topology can be done in linear time using the sum-product algorithm \cite{kschischang2001factor}. Building a linked Bayesian network involves building the Bayesian networks of each relation in a particular order. Indeed, in our example, we first have to build the Bayesian networks of the \textcolor{teal}{customers} and \textcolor{orange}{shops} relations in order to determine the roots that are to be included in the Bayesian network of the \textcolor{violet}{purchases} relation. To build the \textcolor{violet}{purchases} Bayesian network, we first have to join the root attributes (i.e.,\ $nationality$ and $name$) of the first two Bayesian networks (i.e.,\ \textcolor{teal}{customers} and \textcolor{orange}{shops}) with the \textcolor{violet}{purchases} relation. Naturally, performing joins incurs an added computational cost. However, we argue that joins are unavoidable if one is to capture attribute value dependencies across relations. Indeed, if joins are disallowed whatsoever, then there is basically no hope of measuring dependencies between attributes of different relations. Our methodology requires performing one left-join per primary/foreign key relationship, whilst only requiring to include one attribute per join, which is as cost-effective as possible.

The specifics of the procedure we used to build the linked Bayesian network are given in algorithm \ref{algo:build}. We assume the algorithm is given a set of relations. In addition, the algorithm is provided with the set of primary/foreign key relationships in the database (e.g., \textcolor{violet}{purchases} has a foreign key that references \textcolor{teal}{customers}' primary key and another that references \textcolor{orange}{shops}'s primary key). This set of primary/foreign key relationships can easily be extracted from any database's metadata. The idea is to go through the set of relations and check if the Bayesian networks of the dependent relations have been built. In this implementation a \texttt{while} loop is used to go through the relations in their topological order, from bottom to top. The Bayesian networks are built using the $BuildBN$ function, which was presented in \cite{halford2019approach}. The $BuildBN$ function works in three steps: \begin{enumerate*}\item Build a fully-connected, undirected weighted graph, where each node is an attribute and each vertex's weight is the \emph{mutual information} between two attributes. \item Find the \emph{maximum spanning tree} (MST) of the graph. \item Orient the MST in order to obtain a tree by choosing a root.\end{enumerate*}

The $BuildBN$ function produces a Bayesian network with a tree topology called a \emph{Chow-Liu tree} \cite{chow1968approximating}. This tree has the property of being the tree which stores the maximum amount of information out of all the legal trees. In our algorithm, the first pass of the \texttt{while} loop will build the Bayesian networks of the relations that have no dependencies whatsoever (e.g., those who's primary key isn't referenced by any foreign key). The next pass will build the Bayesian networks of the relations that contain primary keys referenced by the foreign keys of the relations covered in the first pass. The algorithm will necessarily terminate once each relation has an associated Bayesian network; it will take as many steps as there are relations in the database.

\begin{algorithm}[H]
  \caption{Linked Bayesian networks construction}
  \begin{algorithmic}[1]
    \Function{BuildLinkedBN}{$relations, relationships$}
    
     \Let{$lbn$}{$\{\}$}  
     \Let{$built$}{$\{\}$}  \Comment{Records which relations have been processed}
     
     \While{$\card{lbn} < \card{relations}$}
        \Let{$queue$}{$relations \setminus built$}   \Comment{Relations which don't have a BN}
            
        \ForEach{$relation \in queue$}
        
            \If{$relationships[relation] \setminus built = \varnothing$}
                \ForEach{$child \in relationships[relation]$}
                    \Let{$relation$}{$relation \bowtie child.root$}
                \EndFor
            \EndIf
            
            \Let{$lbn$}{$lbn \cup BuildBN(relation)$}
            \Let{$built$}{$built \cup relation$}
            
        \EndFor
        
     \EndWhile
     
     \State \Return{$lbn$}
    \EndFunction
     
  \end{algorithmic}
  \label{algo:build}
\end{algorithm}

Note that we can potentially use parallelism to speed-up the execution of algorithm \ref{algo:build}. Indeed, by using a priority queue and a worker pool, we can spawn processes in parallel to build the networks in the correct order. However, we consider this an implementation detail and did not take the time to implement it in our benchmark. Furthermore, this would have skewed our comparison with other methods. A linked Bayesian network doesn't require much more additional space in with respect to the method from \cite{halford2019approach}. Indeed, a linked Bayesian network is nothing more than a set of separate Bayesian networks where some of the attributes are duplicated in two related networks. Once a linked Bayesian network has been built, it can be used to produce selectivity estimates. That is, given a linked Bayesian network, we want to be able to estimate the selectivity of an arbitrary SQL query. An efficient algorithm is required to perform so-called inference when many attributes are involved, which is the topic of the following subsection.

\subsection{Selectivity estimation} \label{querying}.

The algorithm for producing selectivity estimates using linked Bayesian networks is based on the selectivity estimation algorithm proposed in \cite{halford2019approach}. The key insight is that we can fuse linked Bayesian networks into a single Bayesian network. Indeed, in our building process we have to make sure to include the root attribute of each relation's Bayesian network into its parent Bayesian's network. This allows to link each pair of adjacent Bayesian networks together via their shared attribute. In figure \ref{fig:linked-network-topology}, these implicit links are represented with dotted lines. The \textcolor{violet}{purchases} and \textcolor{teal}{customers} relation have in common the $nationality$ attribute, whereas the \textcolor{orange}{shops} and \textcolor{violet}{purchases} relations have in common the $name$ attribute. The resulting ``stiched'' network is necessarily a tree because each individual Bayesian network is a tree and each shared attribute is located at the root of each child network.

\begin{algorithm}[H]
  \caption{Selectivity estimation using a linked Bayesian network}
  \begin{algorithmic}[1]
    \Function{InferSelectivity}{$lbn, query$}
    
     \Let{$relations$}{$ExtractRelations(query)$} 
     \Let{$relevant$}{$PruneLinkedBN(lbn, relations)$}
     \Let{$linked$}{$LinkNetworks(relevant)$}
     \Let{$selectivity$}{$ApplySumProduct(linked)$}
     
     \State \Return{$selectivity$}
    \EndFunction
     
  \end{algorithmic}
  \label{algo:infer}
\end{algorithm}

\begin{figure}[H]
  \centering
        \begin{tikzpicture}[thick, scale=\textwidth, every node/.style={ellipse, draw, align=center},node distance=1cm and 0.2cm]

            \node [draw=violet] (nationality) {Nationality};
            \node [below left = of nationality, draw=teal] (hair) {Hair};
            \node [left = of hair, draw=teal] (salary) {Salary};
            \node [below right = of nationality, xshift=-1cm, draw=violet] (dow) {Day of week};
            \node [right = of dow, draw=violet] (name) {Name};
            
            \node [below left = of name, draw=orange] (city) {City};
            \node [below right = of name, draw=orange] (size) {Size};
            
            \draw [->] (nationality) -- (dow);
            \draw [->] (nationality) -- (name);
            \draw [->] (nationality) -- (hair);
            \draw [->] (nationality) -- (salary);
            
            \draw [->] (name) -- (city);
            \draw [->] (name) -- (size);
        
        \end{tikzpicture}
\caption{Unrolled version of figure \ref{fig:linked-network-topology}}
\label{fig:unrolled-network-topology}
\end{figure}
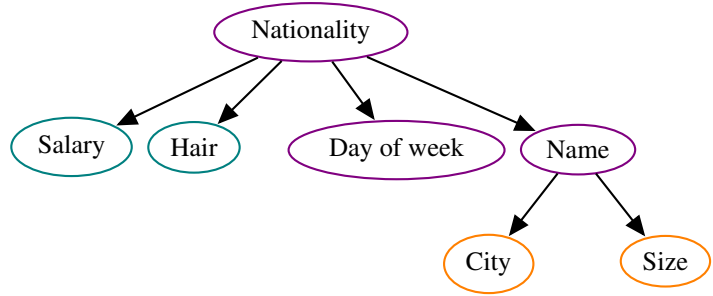

The pseudocode for producing selectivity estimates is given in algorithm \ref{algo:infer}. The first step of the selectivity estimation algorithm is to identify which relations are involved in a given query. Indeed each SQL query will usually involve a subset of relations, and thus we only need to use the Bayesian networks that pertain to said subset. The $PruneLinkedBN$ thus takes care of removing the unnecessary Bayesian networks from the entire set of Bayesian networks. Naturally, in practice, and depending on implementation details, this may involve simply loading in memory the necessary Bayesian networks. In any case, the next step is to connect the networks into a single one. This necessitates looping over the Bayesian networks in topological order -- in the same exact fashion as algorithm \ref{algo:build} -- and linking them along the way. Linking two Bayesian networks together simply involves replacing the attribute they have in common with the child Bayesian network. For instance, in figure \ref{fig:linked-network-topology}, the $nationality$ attribute from the \textcolor{violet}{purchases} Bayesian network will be replaced by the \textcolor{teal}{customers} Bayesian network. This is because we are interested in the distribution of the attributes after the join, not before. The resulting tree thus approximates the distribution of attribute values inside the ($customers \bowtie purchases \bowtie shops$) join instead of estimating selectivities inside each relation independently, as is done in textbook cost models. The result of this linking process is exemplified in figure \ref{fig:unrolled-network-topology}, which shows the unrolled version of the linked Bayesian network shown in figure \ref{fig:linked-network-topology}. Finally, once the Bayesian networks have been linked together, the sum-product algorithm \cite{kschischang2001factor} can be used to output the desired selectivity. In fact, this final step is exactly the same as the one described in section 3.3 of \cite{halford2019approach}.

Our method for estimating selectivities is very efficient. The main reason is because we only to apply the sum-product algorithm once, whereas \cite{halford2019approach} has to apply once per relation involved in the query at hand. This difference is made clear when comparing equations \ref{infer-ibn} and \ref{infer-lbn}. Furthermore, the sum-product algorithm is much more efficient in the case of trees than the clique tree algorithm from \cite{tzoumas2011lightweight}. We confirm these insights in the benchmarks section.

\subsection{Including more than just the roots} \label{subsec:more-than-roots}

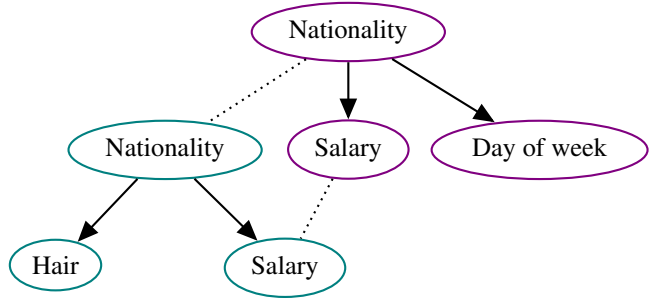
\begin{figure}[H]
  \centering
        \begin{tikzpicture}[thick, every node/.style={ellipse, draw, align=center},node distance=1cm and 0.1cm]

            \node [draw=violet] (p_nationality) {Nationality};
            \node [below = of p_nationality, yshift=0.24cm, draw=violet] (p_salary) {Salary};
            \node [below right = of p_nationality, xshift=0.5cm, draw=violet] (dow) {Day of week};
            
            \node [below left = of p_nationality, xshift=-0.5cm, draw=teal] (nationality) {Nationality};
            \node [below left = of nationality, draw=teal] (hair) {Hair};
            \node [below right = of nationality, draw=teal] (salary) {Salary};
            
            \draw [->] (p_nationality) -- (dow);
            \draw [->] (p_nationality) -- (p_salary);

            \draw [dotted] (p_nationality) -- (nationality);
            \draw [dotted] (p_salary) -- (salary);
            
            \draw [->] (nationality) -- (hair);
            \draw [->] (nationality) -- (salary);
        
        \end{tikzpicture}
\caption{Linked Bayesian network of \textcolor{teal}{customers} and \textcolor{violet}{purchases}}
\label{fig:super-linked-network-topology}
\end{figure}

Our model assumes that the dependencies between attribute values within a relation are preserved when a join occurs. Indeed we assume that tuples are uniformly distributed inside a join \emph{given} each value in the root attribute. One may wonder why we have to stop at the root. Indeed, it turns out that we can include more attributes in addition to the root of each child Bayesian network when building a parent Bayesian network. For example, consider the linked Bayesian network shown in figure \ref{fig:super-linked-network-topology}. In this configuration we include the $salary$ attribute as well as the $nationality$ attribute in the Bayesian network of the \textcolor{violet}{purchases} relation. By doing so we obtain a new conditional distribution $P(salary | nationality)$ which tells us the dependence between \emph{salary} and \emph{nationality} after \textcolor{teal}{customers} has been joined with \textcolor{violet}{purchases}.

The linked Bayesian network shown in \ref{fig:super-linked-network-topology} is valid because we can unroll it in order to obtain a single tree, just as we did earlier on when we only included the \emph{nationality} attribute. However, the $salary$ attribute can be included in the \textcolor{violet}{purchases} Bayesian network only because of the fact that the $nationality$ attribute is included as well. Indeed, if the $nationality$ attribute was not included, then linking \textcolor{teal}{customers} and \textcolor{violet}{purchases} together would have resulted in a Bayesian network which would not necessarily be a tree. In this case, we would not be able to compute $\cp{salary}{nationality}$ in  \textcolor{violet}{purchases}'s Bayesian network. In other words, a node can be included in a parent Bayesian network only if all of its conditioning attributes are included as well. Assuming a child Bayesian network has $n$ nodes, then we can include a number $k \in \{0, \dots, n\}$ of its nodes in the parent Bayesian network. If $k = 0$, then we simply keep each Bayesian network separate, which brings us back to the methodology from \cite{halford2019approach}. If $k = 1$, then we only include the root of each child Bayesian network, which is the case we have discussed up to now. If $k = n$, then we will include all the child's attributes in the parent BN, which is somewhat similar to the global methods presented in \cite{getoor2001selectivity} and \cite{tzoumas2011lightweight}. On the one hand, increasing $k$ will produce larger parent Bayesian networks that capture more attribute value dependencies but also incur a higher computational cost. On the other hand, lower values of $k$ will necessitate less computation but will assume more strongly that dependencies are preserved through joins. The $k$ parameter is thus a practical parameter for compromising between selectivity estimation accuracy and computational requirements. Notice that different values of $k$ can be used for each pair of relations. For instance, we might want to increase $k$ if we notice that the cost model makes very bad estimates for a certain relation. This can be decided upon as deemed fit, be it manually or via automated DBA \cite{van2017automatic}.

\subsection{Summary}

The method we propose attempts to generalise existing selectivity estimation methods based on Bayesian networks. Following the methodology from \cite{halford2019approach}, we build one Bayesian network per relation using Chow-Liu trees. The only difference is that we include a set of attributes from the child relations into the Bayesian network associated with each parent relation. The set of included attributes depends on a chosen parameter $k$ and the structure of each child relation's Bayesian network. Many distributions can be obtained for free because of the fact that each Bayesian network is a tree in which the root attribute conditions the rest of the attributes. This requires assuming that attribute value dependencies are preserved through joins. This assumption, although not always necessarily true, is much softer than the join uniformity as well as the attribute value independence assumptions. The resulting Bayesian networks are thus able to capture attribute value dependencies across relations, as well as inside individual relations. Although our method requires performing joins offline, we argue that joins are unavoidable if one is to capture any cross-relation dependency whatsoever. The major benefit of our method is that it only requires including a single attribute per join, and yet it brings a great deal of information for free through transitivity thanks to our newly introduced assumption. Moreover, our method can still benefit from the efficient selectivity estimation procedure presented in \cite{halford2019approach} because of the preserved tree structure. Finally, our method is able to generalise existing methods based on Bayesian networks through a single parameter which determines the amount of dependency to measure between the attributes of relations that share a primary/foreign key relationship. 

\section{Evaluation}

\subsection{Experimental setup}

We evaluate our proposal on an extensive workload derived from the JOB benchmark \cite{leis2015good}. The JOB benchmark consists of 113 SQL queries, along with an accompanying dataset extracted from the IMDb website. The dataset consists of non-synthetic data, whereas other benchmarks such as TPC-DS \cite{poess2002tpc} are based on synthetic data. The dataset is challenging because it contains skewed distributions and exhibits many correlations between attributes, both across and inside relations. The JOB benchmark is now an established and reliable standard for evaluating and comparing cost models. The dataset and the queries are publicly available \footnote{JOB dataset and queries: \href{https://github.com/gregrahn/join-order-benchmark/}{https://github.com/gregrahn/join-order-benchmark/}}. In addition, we have made a Docker image available for easing future endeavours in the field \footnote{Docker image: \href{https://github.com/MaxHalford/postgres-job-docker}{https://github.com/MaxHalford/postgres-job-docker}}, as well as code used in our experiments \footnote{Method source code: \href{https://github.com/MaxHalford/tldks-2020}{https://github.com/MaxHalford/tldks-2020}}.

During the query optimisation phase, the cost model has to estimate the selectivity of each query execution plan (QEP) enumerated by the query optimiser. Query optimisers usually build QEPs in a bottom-up fashion \cite{chaudhuri1998overview}. Initially, the cost model will have to estimate selectivities for simple QEPs that involve a single relation. It will then be asked to estimate selectivities for larger QEPs involving multiple joins and predicates. We decided to mimic this situation by enumerating all the possible sub-queries for each of the JOB benchmark's queries, as detailed in \cite{chaudhuri2009exact}. For example, if a query pertains to 4 relations, we will enumerate all the possible sub-queries involving 1, 2, 3, and all 4 relations. We also enumerate through all the combinations of filter conditions. To do so, we represented each query as a graph with each node being an attribute and each edge a join. We then simply had to retrieve all the so-called \emph{induced subgraphs}, which are all the subgraphs that can be made up from a given graph. Each induced subgraph was then converted back to a valid SQL statement. This procedure only takes a few minutes and yields a fairly large amount of queries; indeed a total of 5,122,790 subqueries can be generated for the JOB benchmark's 113 queries. Tables \ref{tab:n-joins} and \ref{tab:n-filters} provide an overview of the contents of our workload. 

\begin{table}[!htb]
    \begin{minipage}{.45\linewidth}
      \caption{Query spread per number of join conditions}
      \label{tab:n-joins}
      \centering
        \begin{tabularx}{0.65\textwidth}{@{}YX@{}}
            Joins & Amount \\
            \hline
            0 & 889 \\
            1-5 & 177,309 \\
            6-10 & 1,175,120 \\
            11-15 & 2,060,614 \\
            16-20 & 1,320,681 \\
            21-25 & 388,177 
        \end{tabularx}
    \end{minipage}%
    \hspace{\fill}
    \begin{minipage}{.45\linewidth}
        \caption{Query spread per number of filter conditions}
         \label{tab:n-filters}
      \centering
        \begin{tabularx}{0.65\textwidth}{@{}YX@{}}
            Filters & Amount \\
            \hline 
            1 &    261,440 \\
            2 &    763,392 \\
            3 &   1,301,840 \\
            4 &   1,380,329 \\
            5 &    923,481 \\
            6 &    384,285 \\
            7 &     94,855 \\
            8 &     12,496 \\
            9 &       672
        \end{tabularx}
    \end{minipage} 
\end{table}

The general goal of our experiments is to detail the pros and cons of our method with respect to the textbook approach from \cite{selinger1979access} and some state-of-the-art methods that we were able to implement. Most industrial databases still resort to using textbook approaches, which are thus important to be compared with. Specifically our experiments solely focus on the selectivity estimation module, not on the final query execution time. We assume that improving the selectivity estimates will necessarily have a beneficial impact on the accuracy of the cost model and thus on the query execution time. Naturally, the estimation has to remain reasonable. This seems to be a view shared by many in the query optimisation community \cite{leis2015good}. Indeed, many papers that deal with selectivity estimation, both established and new, do not measure the impact on the final query execution \cite{chen1994adaptive,poosala1996improved,poosala1997selectivity,tzoumas2011lightweight,vengerov2015join,dutt2019selectivity}.

We compared our proposal with a few promising state-of-the-art methods as well as the cardinality estimation module from the PostgreSQL database system. PostgreSQL's cardinality estimation module is a fair baseline as it is a textbook implementation of the decades old ideas from \cite{selinger1979access}. We used version 10.5 of PostgreSQL and did not tinker with the default settings. Additionally, we did not bother with building indexes, as these have no consequence on the selectivity estimation module. A viable selectivity estimation method should be at least as accurate as PostgreSQL, without introducing too much of a computational cost increase. We implemented basic random sampling \cite{olken1986simple}, which consists in executing a given query on a sample of each relation in order to extrapolate a selectivity estimate. Basic random sampling is simple to implement, but isn't suited for queries that involve joins because of the empty-join problem, as explained in section 2. However many sampling methods that take into account the empty-join problem have been proposed. We implemented one such method, namely \emph{correlated sampling} \cite{vengerov2015join}. Correlated sampling works by hashing related primary and foreign keys and discards the tuples of linked relation where the hashes disagree. We also implemented MSCN, which is the deep learning method that is presented in \cite{kipf2019estimating}. Finally we implemented the Bayesian network approach from \cite{tzoumas2011lightweight}. The latter method differs from ours in that it is a global approach that builds one single Bayesian networks over the entire set of relations. Although a global approach is able to capture more correlations than ours, it require more computation. We compared our method with different values for the $k$ parameter presented in section \ref{subsec:more-than-roots}. Note that choosing $k = 0$ is equivalent to using the method from \cite{halford2019approach}. Increasing $k$ is expected to improve the accuracy of the selectivity estimates but deteriorates the computational performance. The $k$ parameter can thus be used to trade between accuracy and computational resources depending on the use case and the constraints of the environment.

\subsection{Selectivity estimation accuracy}

We first measured the accuracy of the selectivity estimates for each method by comparing their estimates with the true selectivity. The true selectivity can be obtained by executing the query and counting the number of tuples in the result. The appropriate metric for such a comparison is called the $q$-error \cite{moerkotte2009preventing,leis2018query}, and is defined as so:

\begin{equation}
q(y, \hat{y}) = \frac{max(y, \hat{y})}{min(y, \hat{y})}
\end{equation}

where $y$ is the true selectivity and $\hat{y}$ is the estimated selectivity. The $q$-error thus simply measures the multiplicative error between the estimate and the truth. The $q$-error has the property of being symmetric, and will thus be the same whether $\hat{y}$ is an underestimation or an overestimation. Moreover the $q$-error is scale agnostic (e.g., $\frac{8}{3} = \frac{24}{9}$), which helps in comparing errors over results with different scales.

\begin{figure}[h]
    \centering
    \scalebox{0.5}{\includegraphics{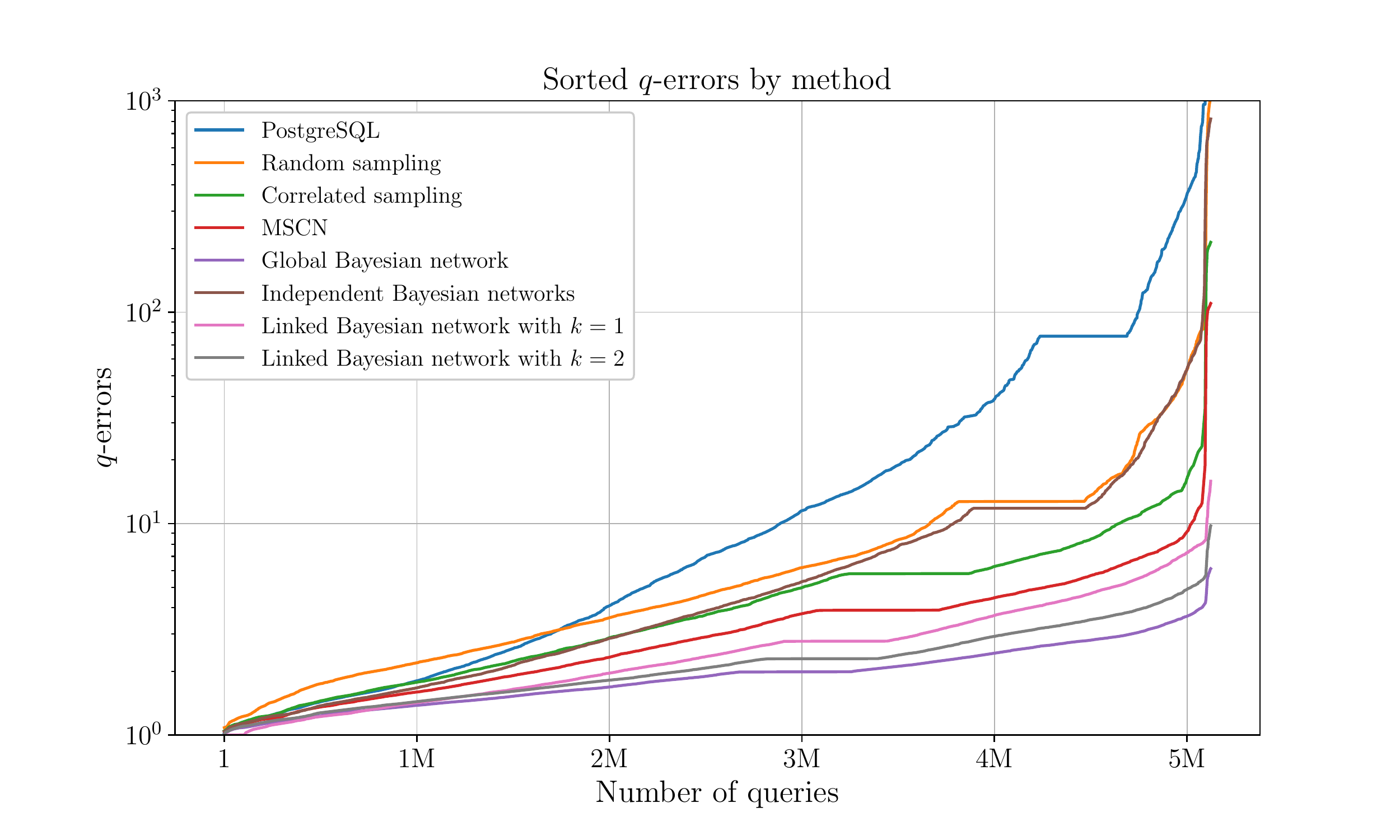}}
    \caption{Sorted $q$-errors for all queries by method on the JOB workload}
    \label{fig:q-errors}
\end{figure}

Figure \ref{fig:q-errors} shows the $q$-errors made by each method for all the queries of the workload derived from the JOB benchmark. The $y$ axis represents the $q$-error associated with each query. Meanwhile the $x$ axis denotes the amount of queries that have less than a given $q$-error. For instance, PostgreSQL managed to estimate the selectivity of two million queries with a $q$-error of less than 10 for each query. The curves thus give us a detailed view into the distribution of the $q$-errors for each method. While the curves seem to exhibit a linear trend, one must note that the scale of the $y$ axis is logarithmic. The figure gives us a global idea of the accuracy of each method in comparison with the others. The mean, maximum, and meaningful quantiles of the $q$-errors are given in table \ref{tab:q-errors}.

\begin{table}[!ht]
  \caption{$q$-error statistics for each method on the JOB workload}
  \label{tab:q-errors}
  \centering
    \begin{tabularx}{\textwidth}{@{}c|YYYYYY@{}}
    &median&90th&95th&99th&max&average\\
    \hline
    PostgreSQL & 7.32 & 77.01 & 185.84 & 707.21 & 10906.17 & 77.01\\
    Sampling & 4.79 & 16.45 & 33.17 & 81.34 & 1018.43 & 12.71\\
    Correlated sampling & 3.83 & 9.63 & 12.63 & 22.72 & 214.1 & 5.79\\
    MSCN & 2.99 & 6.12 & 7.47 & 12.49 & 110.56 & 3.89\\
    Global BN & 1.95 & 2.92 & 3.22 & 4.01 & 7.45 & 1.99\\
    Independent BN & 4.0 & 15.36 & 32.9 & 76.91 & 820.46 & 11.82\\
    Linked BN $k=1$ & 2.41 & 5.03 & 6.15 & 8.07 & 21.09 & 2.79\\
    Linked BN $k=2$ & 2.13 & 3.7 & 4.26 & 5.23 & 12.6 & 2.3\\
    \end{tabularx}
\end{table}

\begin{figure}[h]
    \centering
    \scalebox{0.5}{\includegraphics{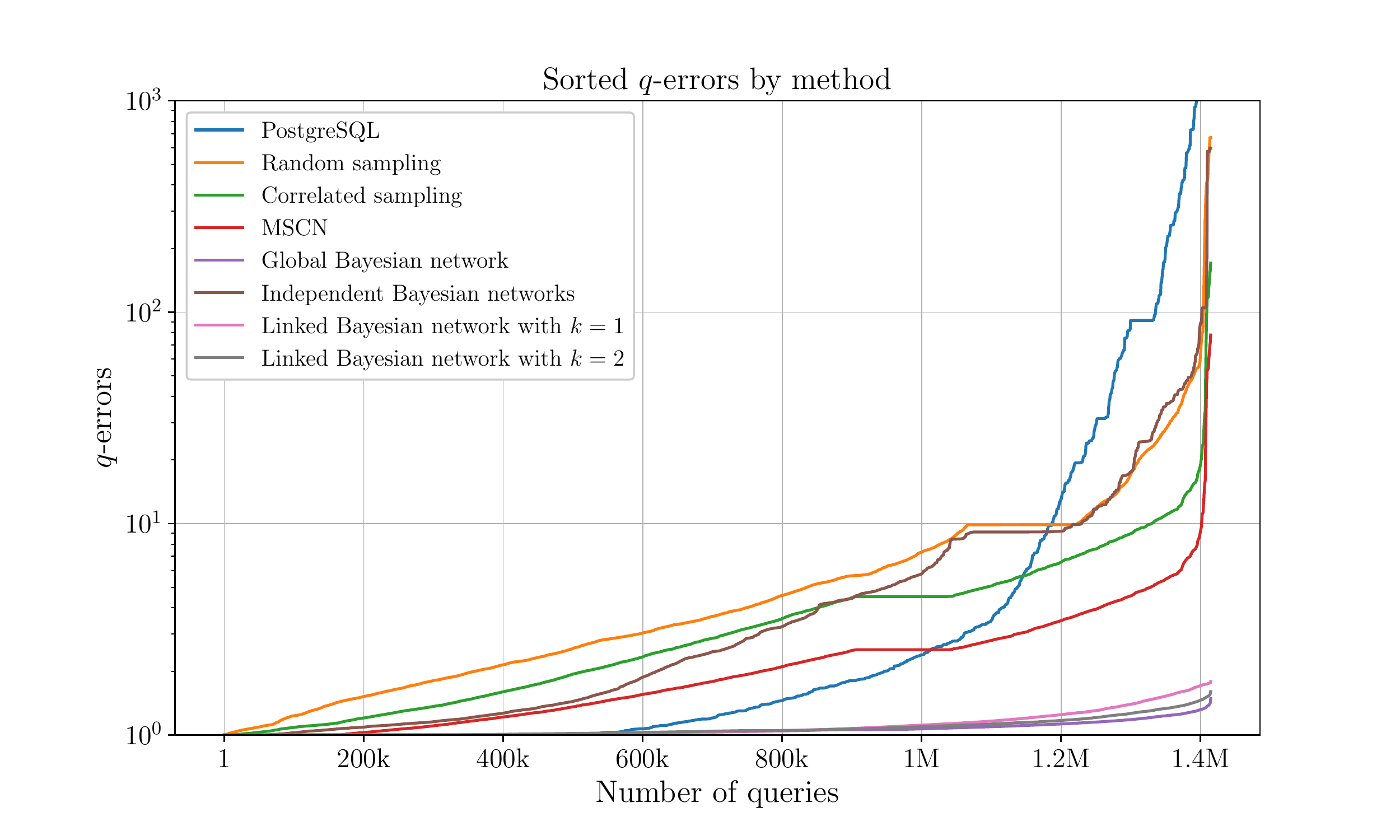}}
    \caption{Sorted $q$-errors for all queries by method on the TPC-DS workload}
    \label{fig:q-errors-tpcds}
\end{figure}

The overall worst method is the cost model used by PostgreSQL. This isn't a surprise, as it assumes total independence between attributes, both within and between relations. It is interesting to notice that the $q$-errors made by PostgreSQL's cost model can be extremely high, sometimes even reaching the tens of thousands. In this case, the query optimiser is nothing short from blind because the selectivity estimates are extremely unreliable. Although this doesn't necessarily mean that the query optimiser will not be able to find a good query execution plan, it does imply that finding a good execution plan would be down to luck \cite{leis2018query}. One may even wonder if estimating a selectivity by picking a random number between 0 and 1 might do better. Using our method with $k$ equal to 0 is equivalent to the methodology proposed by \cite{halford2019approach}. Indeed, if no attributes are shared by the Bayesian networks of each relation, then it is as if we considered attribute value dependencies within each relation but not between relations. As expected, the performance is similar to that of random sampling because both methods capture dependencies within a relation but not between relations. Correlated sampling performs a bit better because it is a join-aware sampling method. However, the rest of the implemented methods seems to be more precise by an order of magnitude. The deep learning method, MSCN, outperforms correlated sampling, but it isn't as performant as the Bayesian networks. However, it can probably reach a better level of performance by tuning some of the many parameters that it exposes. Meanwhile, the method we proposed with $k = 1$ means that we include the root attribute of each child relation within the Bayesian network of each parent relation. This brings to the table the benefits detailed in section 3. If $k = 2$, then an additional attribute from each child relation is included with the Bayesian network of each parent relation. We can see on figure \ref{fig:q-errors-tpcds} that the global accuracy increases with $k$, which is what one would expect. The most accurate method overall is the global Bayesian network presented in \cite{tzoumas2011lightweight}. However, our method with $k = 2$ is not far off. This makes the case that our attribute value dependency preservation assumption is a realistic one.

We have also benchmarked the methods on the TPC-DS benchmark. In contrast to the IMDb dataset used in the JOB benchmark, the TPC-DS dataset is synthetic. By nature, it contains less attribute dependencies than would be expected in a realistic use case. The TPC-DS dataset is therefore less realistic than the JOB benchmark. To produce a workload as we did for the JOB benchmark, we have taken the 30 first queries that are provided with the TPC-DS dataset and have generated all possible sub-queries. This led to a total 1,414,593 queries. The amount of joins went from 2 to 15. The overall results are shown in table \ref{tab:q-errors-postgres}. As expected, the $q$-errors for the TPC-DS benchmark are better across the board because the dataset exhibits less correlations between attributes. Nonetheless, the rankings between the methods remains somewhat the same. Our method very slightly outperforms the global Bayesian network, but we believe that this is just an implementation artifact. In any case, our method is much more accurate than any method that assumes independence between attributes of different relations. Even so, a viable selectivity estimation method also has to be able to produce estimates in a very short amount of time, which is a point we will now discuss.

\begin{table}[!ht]
  \caption{$q$-error statistics for each method on the TPC-DS workload}
  \label{tab:q-errors-postgres}
  \centering
    \begin{tabularx}{\textwidth}{@{}c|YYYYYY@{}}
    &median&90th&95th&99th&max&average\\
    \hline
    PostgreSQL           & 1.23 & 43.4 & 138.05 & 1025.49 & 82898.28 & 91.46\\
    Sampling             & 3.69 & 13.39 & 26.34 & 66.83 & 669.58 & 9.87\\
    Correlated sampling  & 2.89 & 8.24 & 10.63 & 19.32 & 170.63 & 4.51\\
    MSCN                 & 1.82 & 4.23 & 5.32 & 9.24 & 78.01 & 2.54\\
    Global BN            & 1.03 & 1.17 & 1.23 & 1.33 & 1.49 & 1.06\\
    Independent BN       & 2.5 & 13.59 & 33.87 & 89.22 & 597.0 & 9.12\\
    Linked BN $k=1$      & 1.04 & 1.38 & 1.54 & 1.74 & 1.94 & 1.11\\
    Linked BN $k=2$      & 1.05 & 1.26 & 1.35 & 1.47 & 1.6 & 1.08\\
    \end{tabularx}
\end{table}

\subsection{Inference time}

Naturally, we next sought to measure how fast each method was at producing selectivity estimates. In a high throughput environment, the query optimiser isn't allowed to spend much time searching for an efficient QEP. In addition to using the cost model, the query optimiser also has to enumerate potential query execution plans and pick one of them \cite{chaudhuri1998overview}. Thus, only a fraction of the short amount of time allocated to the query optimiser can actually be consumed by the cost model. This means that any viable selectivity estimation has to be extremely efficient, and is probably the main reason why current cost models are kept simple. We call the amount of time necessary to produce a selectivity estimate the \emph{inference time}. During our experiments we recorded the inference time for each query and for each model. The results shown in table \ref{tab:inference-time} show the average inference time for each method, aggregated by the number of joins present in each query.

\begin{table}[!ht]
  \caption{Average inference time in milliseconds for each method with respect to the number of joins on the JOB workload}
  \label{tab:inference-time}
  \centering
    \begin{tabularx}{\textwidth}{@{}c|YYYY@{}}
    & No joins & 1 join & 2 to 5 joins & 6 joins or more \\
\hline
PostgreSQL    & $2.3 \pm 1.1$ & $2.6 \pm 1.4$ & $3.6 \pm 1.3$ & $8.4 \pm 3.1$ \\ 
Sampling      & $19.6 \pm 5.4$ & $36.2 \pm 6.8$ & $120.2 \pm 5.9$ & $268.4 \pm 8.7$ \\ 
Correlated sampling & $20.4 \pm 4.9$ & $155.7 \pm 3.2$ & $280.6 \pm 7.1$ & $493.4 \pm 9.9$ \\ 
MSCN  & $135.9 \pm 12.1$ & $312.2 \pm 24.4$ & $343.3 \pm 27.4$ & $387.6 \pm 29.2$ \\ 
Global BN  & $84.3 \pm 2.1$ & $116.1 \pm 2.9$ & $145.8 \pm 4.4$ & $236.1 \pm 3.8$ \\ 
Independent BN       & $8.3 \pm 1.8$ & $10.9 \pm 1.3$ & $12.6 \pm 2.4$ & $12.1 \pm 3.2$ \\ 
Linked BN $k=1$       & $9.5 \pm 1.9$ & $12.8 \pm 1.6$ & $14.1 \pm 2.8$ & $15.2 \pm 3.4$ \\ 
Linked BN $k=2$       & $10.1 \pm 1.4$ & $12.9 \pm 1.5$ & $14.3 \pm 2.1$ & $16.4 \pm 2.9$
    \end{tabularx}
\end{table}

It is important to mention that the inference time measured for PostgreSQL is simply the time it takes the database to execute the \texttt{ANALYZE} statement for each query. This thus includes the optimisation time, on top of the time spent at estimating selectivities. Even though they are already by far the best, the numbers displayed in our benchmark for PostgreSQL are pessimistic and are expected to be much lower in practice. It is also worth mentioning that we implemented the rest of the methods in Python, which is an interpreted language and thus slower than compiled languages such as C, in which PostgreSQL is written. If these methods were implemented in optimised C they would naturally be much faster. However, what matters here is the relative differences between each method, not the absolute ones.

We can clearly see from the results in table \ref{tab:inference-time} that the global Bayesian network loses in speed what it gains in accuracy. This is because it uses a complex inference method called the \textit{clique-tree algorithm}, which is the standard approach for Bayesian networks with arbitrary topologies. Although it is the most accurate method, it is much slower than our method, regardless of the $k$ parameter we use. What's more, the inference time of our method doesn't increase dramatically when the number of joins increases. This is due to the fact that we use a lightweight inference algorithm called \emph{variable elimination} \cite{cowell2006probabilistic} also used by \cite{halford2019approach}. The inference algorithm scales well because we are able to merge the Bayesian networks of each relation into a single tree. We can also see that correlated sampling is relatively slow method, although its accuracy is competitive as shown in the previous subsection. MSCN is the slowest method overall in our benchmark. This may be attributed to the fact that we implemented it from scratch because no implementation was provided by its authors, and therefore do not have the insights that they might have. We argue that even though our method is not as accurate as the method proposed by \cite{tzoumas2011lightweight}, it is much faster and is thus more likely to be used in practice. Naturally, we also have to take into account the amount of time it requires to build our method, as well as how much storage space it requires.

\subsection{Construction time and space}

The cost model uses metadata that is typically obtained when the database isn't being used. This is done in order not to compute it in real time during the query optimisation phase. This metadata has to be refreshed every so often in order for the cost model to use relevant figures. Typically, the metadata has to be refreshed when the underlying data distributions change significantly. For instance, if attributes become correlated when new data is inserted, then the metadata has to be refreshed to take this into account. Therefore, the amount of time it takes to collect the necessary information is rather important, as ideally we would like to refresh the metadata as often as possible. Additionally, any viable selectivity estimation method crucially has to make do with a little amount of storage space. Indeed, spatial complexity is a major reason why most methods proposed in the literature are not being used in practice. These two computational requirements highlight the dilemma that cost models have to face: they have to be accurate whilst running with a very low footprint. Most multidimensional methods that have been proposed are utterly useless when it comes to their performance in this regard.

\begin{table}[!ht]
  \caption{Computational requirements of the construction phase per method on the JOB workload}
\label{tab:complexity}
  \centering
    \begin{tabularx}{0.8\textwidth}{@{}c|YY@{}}
              & Construction time & Storage size \\
\hline
PostgreSQL    & 5 seconds        & 12KB         \\ 
Sampling      & 7 seconds        & 276MB        \\ 
Correlated sampling & 32 seconds       & 293MB        \\ 
MSCN  & 15 minutes 8 seconds     & 37MB        \\ 
Global BN  & 24 minutes 45 seconds     & 429KB        \\ 
Independent BN       & 55 seconds       & 217KB        \\ 
Linked BN $k=1$       & 2 minutes 3 seconds      & 322KB        \\ 
Linked BN $k=2$       & 2 minutes 8 seconds      & 464KB
    \end{tabularx}
\end{table}

Table \ref{tab:complexity} summarises the computational requirements of the methods we compared. The results explain why PostgreSQL -- and most database engines for that matter -- stick to using simplistic methods. Indeed, in our measurements PostgreSQL is both the fastest method as well the lightest one. PostgreSQL's cost model makes many simplifying assumptions and thus only has to build and store one-dimensional histograms, which can be done extremely rapidly. The sampling methods are quite fast in comparison with the methods based on Bayesian networks. This isn't surprising, as they only require sampling the relations and then storing the samples. Indeed, most of the building time involves persisting the samples on the disk. On the other hand, sampling methods require a relatively large amount of space because they do not apply any summarising whatsoever (note that their storage size are given in terms of megabytes, not kilobytes). Correlated sampling takes more time than basic sampling because it has to scan primary and foreign keys in order to avoid the empty join problem. MSCN construction time is moderate, and naturally depends on the amount of data it is trained on. In this case we trained it for 20 epochs of stochastic gradient descent.

All of the methods based on Bayesian networks take more time to build than the two sampling methods, which is as expected. They make up in storage requirements and in inference time. The global Bayesian network takes a very large amount of time to build, which is in accordance with the results from \cite{tzoumas2011lightweight}. In comparison, our method is much faster. This is a logical consequence of the fact that we only build one Bayesian network per relation. Additionally, each Bayesian network has a tree topology, which means that each conditional probability distribution we need to store is a two-way table. The sudden jump in building time between $k = 0$ and $k = 1$ is due to the need to compute joins when $k > 0$. However, note that the jump is much smaller between $k = 1$ and $k = 2$. The reason is that the joins don't have to be repeated for each additional attribute included in every parent Bayesian network.

\section{Conclusion}

During the query optimisation phase, a cost model is invoked by the query optimiser to estimate the cost of query execution plans. In this context, the selectivity of operators is a crucial input to the cost model \cite{leis2015good}. Inaccurate selectivity estimates lead to bad cost estimates which in turn have a negative impact on the overall running time of a query. Moreover, errors in selectivity estimation grow exponentially throughout a query execution plan \cite{ioannidis1991propagation}. Selectivity estimation is still an open research problem, even though many proposals have been made. This is down to the fact that the requirements in terms of computational resources are extremely tight, and one thus has to compromise between accuracy and efficiency.

Our method is based on Bayesian networks, which are a promising way to solve the aforementioned compromise. Although the use of Bayesian networks for selectivity estimation isn't new, previous propositions entail a prohibitive building cost and inference time. In order to address these issues, we extend the work of \cite{halford2019approach} to include the measurement of dependencies between attributes of different relations. We show how we can soften the relational independence assumption without requiring an inordinate amount of computational resources. We validate our method by comparing it with other methods on an extensive workload derived from the JOB \cite{leis2015good} and the TPC-DS \cite{poess2002tpc} benchmarks. Our results show that our method is only slightly less accurate than the global Bayesian network from \cite{tzoumas2011lightweight}, whilst being an order of magnitude less costly to build and execute. Additionally, our method is more accurate than join-aware sampling, whilst requiring significantly less storage and computational requirements. In comparison with other methods which make more simplifying assumptions, our method is notably more accurate, whilst offering very reasonable guarantees in terms of computational time and space. In future work, we wish to extend our method to accommodate for specific operators such as \texttt{GROUP BY}s, as well as verify the benefits of our method in terms of overall query response time as perceived by a query issuer.

\bibliographystyle{unsrt}  
%\bibliography{template}  %%% Remove comment to use the external .bib file (using bibtex).


\begin{thebibliography}{10}

\bibitem{ioannidis1996query}
Yannis~E Ioannidis.
\newblock Query optimization.
\newblock {\em ACM Computing Surveys (CSUR)}, 28(1):121--123, 1996.

\bibitem{leis2015good}
Viktor Leis, Andrey Gubichev, Atanas Mirchev, Peter Boncz, Alfons Kemper, and
  Thomas Neumann.
\newblock How good are query optimizers, really?
\newblock {\em Proceedings of the VLDB Endowment}, 9(3):204--215, 2015.

\bibitem{ioannidis1991propagation}
Yannis~E Ioannidis and Stavros Christodoulakis.
\newblock {\em On the propagation of errors in the size of join results},
  volume~20.
\newblock ACM, 1991.

\bibitem{yin2018sla}
Shaoyi Yin, Abdelkader Hameurlain, and Franck Morvan.
\newblock Sla definition for multi-tenant dbms and its impact on query
  optimization.
\newblock {\em IEEE Transactions on Knowledge and Data Engineering}, 2018.

\bibitem{wu2013predicting}
Wentao Wu, Yun Chi, Shenghuo Zhu, Junichi Tatemura, Hakan Hacig{\"u}m{\"u}s,
  and Jeffrey~F Naughton.
\newblock Predicting query execution time: Are optimizer cost models really
  unusable?
\newblock In {\em 2013 IEEE 29th International Conference on Data Engineering
  (ICDE)}, pages 1081--1092. IEEE, 2013.

\bibitem{selinger1979access}
P~Griffiths Selinger, Morton~M Astrahan, Donald~D Chamberlin, Raymond~A Lorie,
  and Thomas~G Price.
\newblock Access path selection in a relational database management system.
\newblock In {\em Proceedings of the 1979 ACM SIGMOD international conference
  on Management of data}, pages 23--34. ACM, 1979.

\bibitem{ioannidis2003history}
Yannis Ioannidis.
\newblock -the history of histograms (abridged).
\newblock In {\em Proceedings 2003 VLDB Conference}, pages 19--30. Elsevier,
  2003.

\bibitem{muralikrishna1988equi}
M~Muralikrishna and David~J DeWitt.
\newblock Equi-depth multidimensional histograms.
\newblock In {\em ACM SIGMOD Record}, volume~17, pages 28--36. ACM, 1988.

\bibitem{chen2017two}
Yu~Chen and Ke~Yi.
\newblock Two-level sampling for join size estimation.
\newblock In {\em Proceedings of the 2017 ACM International Conference on
  Management of Data}, pages 759--774. ACM, 2017.

\bibitem{chaudhuri1999random}
Surajit Chaudhuri, Rajeev Motwani, and Vivek Narasayya.
\newblock On random sampling over joins.
\newblock In {\em ACM SIGMOD Record}, volume~28, pages 263--274. ACM, 1999.

\bibitem{stillger2001leo}
Michael Stillger, Guy~M Lohman, Volker Markl, and Mokhtar Kandil.
\newblock Leo-db2's learning optimizer.
\newblock In {\em VLDB}, volume~1, pages 19--28, 2001.

\bibitem{halford2019approach}
Max Halford, Philippe Saint-Pierre, and Franck Morvan.
\newblock An approach based on bayesian networks for query selectivity
  estimation.
\newblock In {\em International Conference on Database Systems for Advanced
  Applications}, pages 3--19. Springer, 2019.

\bibitem{jensen1996introduction}
Finn~V Jensen et~al.
\newblock {\em An introduction to Bayesian networks}, volume 210.
\newblock UCL press London, 1996.

\bibitem{cooper1990computational}
Gregory~F Cooper.
\newblock The computational complexity of probabilistic inference using
  bayesian belief networks.
\newblock {\em Artificial intelligence}, 42(2-3):393--405, 1990.

\bibitem{getoor2001selectivity}
Lise Getoor, Benjamin Taskar, and Daphne Koller.
\newblock Selectivity estimation using probabilistic models.
\newblock In {\em ACM SIGMOD Record}, volume~30, pages 461--472. ACM, 2001.

\bibitem{tzoumas2011lightweight}
Kostas Tzoumas, Amol Deshpande, and Christian~S Jensen.
\newblock Lightweight graphical models for selectivity estimation without
  independence assumptions.
\newblock {\em Proceedings of the VLDB Endowment}, 4(11):852--863, 2011.

\bibitem{poess2002tpc}
Meikel Poess, Bryan Smith, Lubor Kollar, and Paul Larson.
\newblock Tpc-ds, taking decision support benchmarking to the next level.
\newblock In {\em Proceedings of the 2002 ACM SIGMOD international conference
  on Management of data}, pages 582--587, 2002.

\bibitem{leis2018query}
Viktor Leis, Bernhard Radke, Andrey Gubichev, Atanas Mirchev, Peter Boncz,
  Alfons Kemper, and Thomas Neumann.
\newblock Query optimization through the looking glass, and what we found
  running the join order benchmark.
\newblock {\em The VLDB Journal}, pages 1--26, 2018.

\bibitem{kooi1981optimization}
Robert~Philip Kooi.
\newblock The optimization of queries in relational databases.
\newblock 1981.

\bibitem{blohsfeld1999comparison}
Bj{\"o}rn Blohsfeld, Dieter Korus, and Bernhard Seeger.
\newblock A comparison of selectivity estimators for range queries on metric
  attributes.
\newblock In {\em ACM SIGMOD Record}, volume~28, pages 239--250. ACM, 1999.

\bibitem{matias1998wavelet}
Yossi Matias, Jeffrey~Scott Vitter, and Min Wang.
\newblock Wavelet-based histograms for selectivity estimation.
\newblock In {\em ACM SIGMoD Record}, volume~27, pages 448--459. ACM, 1998.

\bibitem{poosala1997selectivity}
Viswanath Poosala and Yannis~E Ioannidis.
\newblock Selectivity estimation without the attribute value independence
  assumption.
\newblock In {\em VLDB}, volume~97, pages 486--495, 1997.

\bibitem{heimel2015self}
Max Heimel, Martin Kiefer, and Volker Markl.
\newblock Self-tuning, gpu-accelerated kernel density models for
  multidimensional selectivity estimation.
\newblock In {\em Proceedings of the 2015 ACM SIGMOD International Conference
  on Management of Data}, pages 1477--1492. ACM, 2015.

\bibitem{bruno2001stholes}
Nicolas Bruno, Surajit Chaudhuri, and Luis Gravano.
\newblock Stholes: a multidimensional workload-aware histogram.
\newblock In {\em Acm Sigmod Record}, volume~30, pages 211--222. ACM, 2001.

\bibitem{vengerov2015join}
David Vengerov, Andre~Cavalheiro Menck, Mohamed Zait, and Sunil~P Chakkappen.
\newblock Join size estimation subject to filter conditions.
\newblock {\em Proceedings of the VLDB Endowment}, 8(12):1530--1541, 2015.

\bibitem{leis2017cardinality}
Viktor Leis, Bernharde Radke, Andrey Gubichev, Alfons Kemper, and Thomas
  Neumann.
\newblock Cardinality estimation done right: Index-based join sampling.
\newblock In {\em CIDR}, 2017.

\bibitem{acharya1999join}
Swarup Acharya, Phillip~B Gibbons, Viswanath Poosala, and Sridhar Ramaswamy.
\newblock Join synopses for approximate query answering.
\newblock In {\em ACM SIGMOD Record}, volume~28, pages 275--286. ACM, 1999.

\bibitem{markl2007consistent}
Volker Markl, Peter~J Haas, Marcel Kutsch, Nimrod Megiddo, Utkarsh Srivastava,
  and Tam~Minh Tran.
\newblock Consistent selectivity estimation via maximum entropy.
\newblock {\em The VLDB journal}, 16(1):55--76, 2007.

\bibitem{muller2018improved}
Magnus M{\"u}ller, Guido Moerkotte, and Oliver Kolb.
\newblock Improved selectivity estimation by combining knowledge from sampling
  and synopses.
\newblock {\em Proceedings of the VLDB Endowment}, 11(9):1016--1028, 2018.

\bibitem{akdere2012learning}
Mert Akdere, Ugur {\c{C}}etintemel, Matteo Riondato, Eli Upfal, and Stanley~B
  Zdonik.
\newblock Learning-based query performance modeling and prediction.
\newblock In {\em Data Engineering (ICDE), 2012 IEEE 28th International
  Conference on}, pages 390--401. IEEE, 2012.

\bibitem{liu2015cardinality}
Henry Liu, Mingbin Xu, Ziting Yu, Vincent Corvinelli, and Calisto Zuzarte.
\newblock Cardinality estimation using neural networks.
\newblock In {\em Proceedings of the 25th Annual International Conference on
  Computer Science and Software Engineering}, pages 53--59. IBM Corp., 2015.

\bibitem{ivanov2017adaptive}
Oleg Ivanov and Sergey Bartunov.
\newblock Adaptive cardinality estimation.
\newblock {\em arXiv preprint arXiv:1711.08330}, 2017.

\bibitem{kipf2018learned}
Andreas Kipf, Thomas Kipf, Bernhard Radke, Viktor Leis, Peter Boncz, and Alfons
  Kemper.
\newblock Learned cardinalities: Estimating correlated joins with deep
  learning.
\newblock {\em arXiv preprint arXiv:1809.00677}, 2018.

\bibitem{kipf2019estimating}
Andreas Kipf, Dimitri Vorona, Jonas M{\"u}ller, Thomas Kipf, Bernhard Radke,
  Viktor Leis, Peter Boncz, Thomas Neumann, and Alfons Kemper.
\newblock Estimating cardinalities with deep sketches.
\newblock In {\em Proceedings of the 2019 International Conference on
  Management of Data}, pages 1937--1940, 2019.

\bibitem{deshpande2001independence}
Amol Deshpande, Minos Garofalakis, and Rajeev Rastogi.
\newblock Independence is good: Dependency-based histogram synopses for
  high-dimensional data.
\newblock {\em ACM SIGMOD Record}, 30(2):199--210, 2001.

\bibitem{bartlett2017integer}
Mark Bartlett and James Cussens.
\newblock Integer linear programming for the bayesian network structure
  learning problem.
\newblock {\em Artificial Intelligence}, 244:258--271, 2017.

\bibitem{chow1968approximating}
C~Chow and Cong Liu.
\newblock Approximating discrete probability distributions with dependence
  trees.
\newblock {\em IEEE transactions on Information Theory}, 14(3):462--467, 1968.

\bibitem{cowell2006probabilistic}
Robert~G Cowell, Philip Dawid, Steffen~L Lauritzen, and David~J Spiegelhalter.
\newblock {\em Probabilistic networks and expert systems: Exact computational
  methods for Bayesian networks}.
\newblock Springer Science \& Business Media, 2006.

\bibitem{hwang1992steiner}
Frank~K Hwang, Dana~S Richards, and Pawel Winter.
\newblock {\em The Steiner tree problem}, volume~53.
\newblock Elsevier, 1992.

\bibitem{kschischang2001factor}
Frank~R Kschischang, Brendan~J Frey, Hans-Andrea Loeliger, et~al.
\newblock Factor graphs and the sum-product algorithm.
\newblock {\em IEEE Transactions on information theory}, 47(2):498--519, 2001.

\bibitem{van2017automatic}
Dana Van~Aken, Andrew Pavlo, Geoffrey~J Gordon, and Bohan Zhang.
\newblock Automatic database management system tuning through large-scale
  machine learning.
\newblock In {\em Proceedings of the 2017 ACM International Conference on
  Management of Data}, pages 1009--1024. ACM, 2017.

\bibitem{chaudhuri1998overview}
Surajit Chaudhuri.
\newblock An overview of query optimization in relational systems.
\newblock In {\em Proceedings of the seventeenth ACM SIGACT-SIGMOD-SIGART
  symposium on Principles of database systems}, pages 34--43. ACM, 1998.

\bibitem{chaudhuri2009exact}
Surajit Chaudhuri, Vivek Narasayya, and Ravi Ramamurthy.
\newblock Exact cardinality query optimization for optimizer testing.
\newblock {\em Proceedings of the VLDB Endowment}, 2(1):994--1005, 2009.

\bibitem{chen1994adaptive}
Chungmin~Melvin Chen and Nick Roussopoulos.
\newblock {\em Adaptive selectivity estimation using query feedback},
  volume~23.
\newblock ACM, 1994.

\bibitem{poosala1996improved}
Viswanath Poosala, Peter~J Haas, Yannis~E Ioannidis, and Eugene~J Shekita.
\newblock Improved histograms for selectivity estimation of range predicates.
\newblock In {\em ACM Sigmod Record}, volume~25, pages 294--305. ACM, 1996.

\bibitem{dutt2019selectivity}
Anshuman Dutt, Chi Wang, Azade Nazi, Srikanth Kandula, Vivek Narasayya, and
  Surajit Chaudhuri.
\newblock Selectivity estimation for range predicates using lightweight models.
\newblock {\em Proceedings of the VLDB Endowment}, 12(9):1044--1057, 2019.

\bibitem{olken1986simple}
Frank Olken and Doron Rotem.
\newblock Simple random sampling from relational databases.
\newblock 1986.

\bibitem{moerkotte2009preventing}
Guido Moerkotte, Thomas Neumann, and Gabriele Steidl.
\newblock Preventing bad plans by bounding the impact of cardinality estimation
  errors.
\newblock {\em Proceedings of the VLDB Endowment}, 2(1):982--993, 2009.

\bibitem{kruskal1956shortest}
Joseph~B Kruskal.
\newblock On the shortest spanning subtree of a graph and the traveling
  salesman problem.
\newblock {\em Proceedings of the American Mathematical society}, 7(1):48--50,
  1956.

\end{thebibliography}

\clearpage
\section{Supplementary materials}

\subsection{Preliminary works}

The following is an unabbreviated version of the subsection on our our preliminary subsection. It contains additional examples that help to get a better understanding, but that were considered too lengthy to be part of the main article.

In \cite{halford2019approach}, we developed a methodology for constructing Bayesian networks to model the distribution of attribute values inside each relation of a database. Once the Bayesian networks are constructed, we used to produce selectivity estimates by converting a logical operator tree at hand into a probabilistic formula of sums and products. In what follows, we will give an overview of Bayesian networks. We will also explain the compromises we made in order to produce a method that is both reasonably accurate as well as efficient.

A Bayesian network is a probabilistic model. As such, it is used for approximating the probability distribution of a dataset. The particularity of a Bayesian network is that it uses a directed acyclic graph (DAG) in order to do so. The graph contains one node per variable, whilst each directed edge represents a conditional dependency between two variables. For instance, if nodes $A$ and $B$ are connected with an edge that points from $A$ to $B$, then this stands for the conditional distribution $P(B \, | \, A)$. A Bayesian network is a product of many such conditional dependencies, which formally is:

\begin{equation}
    P(X_1, \dots, X_n) \simeq \prod_{X_i \in \mathcal{X}} P(X_i \, | \, Parents(X_i))
\end{equation}

The term, $P(X_1, \dots, X_n)$ is the probability distribution over the entire set of attributes $\{X_1, \dots, X_n\}$. Meanwhile, $Parents(X_i)$ stands for the attributes that condition the value of $X_i$. The distribution $P(X_i \, | \, Parents(X_i))$ is thus the conditional distribution of attribute $X_i$'s value. In practice, the full distribution is inordinately large, and is unknown to us. However, the total of the sizes of the conditional distributions $P(X_i \, | \, Parents(X_i))$ is much smaller. Indeed, for discrete attributes, each conditional distribution is a $(p+1)$-way table, where $p$ is the number of parents $\card{Parent(X_i)}$. If an attribute $hair$ is conditioned by a single other attribute $nationality$, then that conditional relationship can be stored in a two-way table, as shown in table \ref{tab:hair-nationality}.

\begin{table}[H]
      \centering
\caption{Conditional distribution $P(hair \, | \, nationality)$}
\label{tab:hair-nationality}
\begin{tabularx}{0.65\textwidth}{@{}YYYY@{}}
           & \textbf{Blond} & \textbf{Brown} & \textbf{Dark} \\ \hline
\textbf{American}     & 0.3  & 0.4  & 0.3   \\ \hline
\textbf{Japanese} & 0.05 & 0.1 & 0.85  \\  \hline
\textbf{Swedish} & 0.7 & 0.2 & 0.1 
\end{tabularx}
\end{table}

Meanwhile, if the $nationality$ attribute is not conditioned by any other attribute, then we can represent with a one-dimensional distribution, which is represented in table \ref{tab:nationality}.

\begin{table}[H]
      \centering
\caption{Distribution $P(nationality)$}
\label{tab:nationality}
\begin{tabularx}{0.5\textwidth}{@{}YYY@{}}
           \textbf{American} & \textbf{Japanese} & \textbf{Swedish} \\ \hline
            0.2  & 0.5  & 0.3
\end{tabularx}
\end{table}

Assume the cost model is asked by the query optimiser to determine the fraction of tuples where $hair$ equals $``Blond"$ and $nationality$ equals $``Swedish"$. This can be obtained by applying Bayes' rule:

\begin{equation}
\begin{split}
    P(hair=Blond, nationality=Swedish) & = P(Blond \, | \, Swedish) \times P(Swedish) \\
    & = 0.7 \times 0.3 \\
    & = 0.21
\end{split}
\end{equation}

Note that we do not have directly access to the distribution of the $hair$ attribute. Indeed we only have the conditional distribution $P(hair \, | \, nationality)$. However, we can obtain $P(hair)$ by marginalising over the $nationality$ attribute. For instance, if we want to obtain the fraction of tuples where $hair$ equals $``Blond"$:

\begin{equation}
\begin{split}
    P(hair=Blond) & = \sum_{nationality} P(Blond, nationality) \\
                  & = \sum_{nationality} P(Blond \, | \, nationality) \times P(nationality) \\
                  & = \underbrace{0.3 \times 0.2}_{P(Blond, American)} + \underbrace{0.05 \times 0.5}_{P(Blond, Japanese)} + \underbrace{0.7 \times 0.3}_{P(Blond, Swedish)} \\
                  & = 0.295
\end{split}
\end{equation}

With a Bayesian network, we are thus able to answer any selectivity estimation problem by converting a logical query into a mathematical formula following standard rules of probability \cite{jensen1996introduction}. Note, however, that a Bayesian network is necessarily an approximation of the full probability distribution because it makes assumptions about the generating process of the data. Finding the right graph structure of a Bayesian network is called \emph{structure learning} \cite{jensen1996introduction}. This is usually done using a scoring function, which estimates the amount of information memorised by a network with a given structure. The time required to run an exhaustive search which maximises the scoring function is super-exponential with the number of variables \cite{cooper1990computational}. Approximate search methods as well as integer programming solutions have been proposed \cite{bartlett2017integer}, but they still require a large amount of time to run and have brittle performance guarantees. In our work in \cite{halford2019approach}, we proposed to use the \emph{Chow-Liu} algorithm \cite{chow1968approximating}. This algorithm has the property of finding the best tree structure where nodes are restricted to have at most one parent. The obtained tree is the best in the sense of maximum likelihood estimation. In other words, it is the tree that memorises the most the given data. This is an important property, because for the purpose of selectivity estimation we are not interested in having a model that generalises well, but rather one that is good at memorising the data that it is shown. This is explained in further details in section 4.1 of \cite{getoor2001selectivity}. In addition to this property, the Chow-Liu algorithm only runs in $\mathcal{O}(p^2)$ time, where $p$ is the number of variables, and is simple to implement. It works by first computing the \emph{mutual information} between each pair of variables, which is defined as so:

\begin{equation}
    MI(X_i, X_j) = \sum_{x_i \in X_i} \sum_{x_j \in X_j} P(x_i, x_j) \times \log(\frac{P(x_i, x_j)}{P(x_i)P(x_j)})
\end{equation}

The mutual information can be seen as the strength of the relation between two variables, whether it be linear or not. The distribution $P(X_i, X_j)$ contains the occurrence counts of each pair $(x_i, x_j)$ in a relation $R$. It can be obtained with a \texttt{SELECT COUNT(*) FROM R GROUP BY $X_i$, $X_j$} statement in \texttt{SQL}. The distributions $P(X_i)$ and $P(X_j)$ can be obtained by marginalising over $P(X_i, X_j)$ with respect to the other attribute. Once the mutual information for each pair of attributes is computed, they are organised into a fully connected weighted graph, as shown in figure \ref{fig:fully-connected}:

\begin{figure}[H]
  \centering
    \begin{tikzpicture}[auto, node distance=0.7cm]
       \begin{scope}
        \node (v1) at (0,2.5) {nationality};
        \node (v2) at (-3,0.5) {country};
        \node (v3) at (3,0.5) {city};
        \node (v4) at (-1.8,-2.5) {eye colour};
        \node (v5) at (1.8,-2.5) {hair colour};
       \end{scope}
       \begin{scope}
        \draw  (v1) edge node{0.55} (v2);
        \draw  (v1) edge node{0.34} (v3);
        \draw  (v1) edge node{0.11} (v4);
        \draw  (v1) edge node{0.59} (v5);
        \draw  (v2) edge node{0.68} (v3);
        \draw  (v2) edge node{0.03} (v4);
        \draw  (v2) edge node{0.25} (v5);
        \draw  (v3) edge node{0.01} (v4);
        \draw  (v3) edge node{0.22} (v5);
        \draw  (v4) edge node{0.10} (v5);
       \end{scope}
    \end{tikzpicture}
    \caption{Mutual information amounts for five attributes}
    \label{fig:fully-connected}
\end{figure}
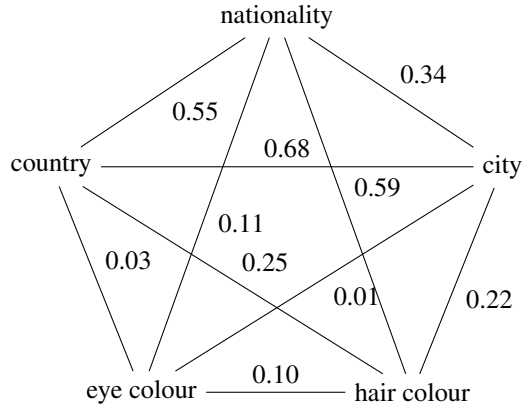

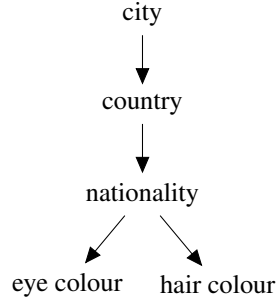
\begin{figure}[H]
  \centering
    \begin{tikzpicture}[auto, node distance=2cm]
       \begin{scope}
        \node (v3) at (0,2.4) {city};
        \node (v2) at (0,1.2) {country};
        \node (v1) at (0,0) {nationality};
        \node (v4) at (-1,-1.2) {eye colour};
        \node (v5) at (1,-1.2) {hair colour};
       \end{scope}
       \begin{scope}
        \draw [->] (v3) edge (v2);
        \draw [->] (v2) edge (v1);
        \draw [->] (v1) edge (v4);
        \draw [->] (v1) edge (v5);
       \end{scope}
    \end{tikzpicture}
    \caption{Maximum spanning tree of figure \ref{fig:fully-connected}}
    \label{fig:mst}
\end{figure}

The next step is to find the \emph{maximum spanning tree} (MST) of the graph, which is the spanning tree whose sum of edge weights is maximal. A spanning tree is a subset of $p-1$ edges that forms a tree. Finding the maximum spanning tree can be done in $\mathcal{O}(p\log(p))$ time, for example, by using Kruskal's algorithm \cite{kruskal1956shortest}. The maximum spanning tree is then turned into a directed graph by choosing a root attribute. The choice of the root attribute does not matter because mutual information is symmetric. The result of applying this procedure to the graph from figure \ref{fig:fully-connected} is shown in figure \ref{fig:mst}.

Once the structure of a Bayesian network has been decided upon, it can be used to answer probabilistic queries. This is usually referred to as \textit{inference}. Inference for Bayesian networks is known to be NP-hard \cite{cooper1990computational}. Exact inference as well as approximate methods have been proposed. The most basic inference algorithm is called the \emph{variable elimination} algorithm \cite{cowell2006probabilistic} and is an exact inference algorithm. It works by explicitly writing the inference equation defined by the Bayesian network's structures, and then moving the sum and product operators around in order to "eliminate" repetitive calculations. In the case of trees, this can be done in linear time, which is the main reason why we constrained our Bayesian networks to have tree topologies in \cite{halford2019approach}. Many other inference algorithms exist; this includes belief propagation (used in \cite{tzoumas2011lightweight}), linear programming, sampling methods, and variational inference. However, our experiments indicated that all of these were much slower than the variable elimination algorithm in the case of trees. Our inference process can further be accelerated by identifying branches of the Bayesian network that do not pertain to a particular query. This identification process is called the \emph{Steiner tree problem} \cite{hwang1992steiner}.

In \cite{halford2019approach}, we proposed a simple method which consists in building one Bayesian network per relation. We used the Bayesian networks to estimate the selectivity of queries inside their respectful relations. On the one hand, this has the benefit of greatly reducing the computational burden in comparison with a single large Bayesian network, as is done in \cite{getoor2001selectivity} and \cite{tzoumas2011lightweight}. On the other hand, it ignores dependencies between attributes of different relations. We will now discuss how we can improve our work from \cite{halford2019approach} in order to capture some dependencies across relations.

\end{document}